\documentclass[traditabstract,article,a4paper]{aa}  
\usepackage{graphicx,txfonts}
\usepackage{natbib}
\usepackage[dvipsnames]{xcolor}
\bibpunct{(}{)}{;}{a}{}{,} % A&A style

\begin{document}

\newcommand{\todo}{\textcolor{purple}}
\newcommand{\todog}{\textcolor{green}}
\newcommand{\mytodo}{\colorbox{orange}}
\newcommand{\ybox}[2][yellow]{
 \colorbox{#1}{\parbox{\dimexpr\linewidth-2\fboxsep}{\strut #2\strut}}%
}
\newcommand{\limebox}[2][yellow]{%
  \colorbox{#1}{\parbox{\dimexpr\linewidth-2\fboxsep}{\strut #2\strut}}%
}
%
%\usepackage{graphicx}
%%%%%%%%%%%%%%%%%%%%%%%%%%%%%%%%%%%%%%%%
%\usepackage{txfonts}
%%%%%%%%%%%%%%%%%%%%%%%%%%%%%%%%%%%%%%%%
%\usepackage[options]{hyperref}
% To add links in your PDF file, use the package "hyperref"
% with options according to your LaTeX or PDFLaTeX drivers.
%
\title{Unveiling the largest structures in the nearby Universe: \\Discovery of the Quipu superstructure}

   \author{Hans B\"ohringer\inst{1,2,3} \and Gayoung Chon\inst{2,3}  \and Joachim Tr\"umper\inst{1}
     Renee C. Kraan-Korteweg\inst{4} \and  Norbert Schartel\inst{5}    }

   \institute{
   Max-Planck-Institut f\"ur Extraterrestrische Physik, Giessenbachstr.~1, 85748 Garching, Germany
        \and     
   Max-Planck-Institut f\"ur Physik, Boltzmannstr.~8, 85748 Garching, Germany
           \and    
   Universit\"ats-Sternwarte M\"unchen, Fakult\"at f\"ur Physik, Ludwig-Maximilians-Universit\"at M\"unchen, Scheinerstr. 1, 81679 M\"unchen, Germany
        \and
 Dept. of Astronomy, University of Cape Town, Privat Bag X3, Rondebosch, 7701, South Africa
        \and                    
ESAC, European Space Agancy, Camino bajo del Castillo, Villanueva de la Ca\~nada, 28692, Spain.
}

   \date{Received ... 2024; accepted ...}

\abstract{
For a precise determination of cosmological parameters we need to understand the effects of the local large-scale structure of the Universe on the measurements. 
They include modifications of the cosmic microwave background, distortions of sky images by large-scale gravitational lensing, and the influence of large-scale streaming motions on measurements of the Hubble constant. 
The streaming motions, for example, originate from mass concentrations with distances up to 250~Mpc. In this paper we provide the first all-sky assessment of the
largest structures at distances between 130 and 250~Mpc and discuss their observational consequences, using X-ray galaxy clusters to map the matter density distribution.
Among the five most prominent superstructures found, the largest has a length longer than $400$~Mpc with an estimated mass of about $2\times10^{17}$~M$_{\odot}$.
This entity, which we named Quipu, is the largest cosmic structure discovered to date.
These superstructures contain about 45\% of the galaxy clusters, 30\% of the galaxies, 25\% of the matter, and occupy a volume fraction of 13\%, thus constituting a major part of the Universe.
The galaxy density is enhanced in the environment of superstructures out to larger distances from the nearest member clusters compared to the outskirts of clusters in the field.
We find superstructures with similar properties in simulations based on $\Lambda$CDM cosmology models.
We show that the superstructures should produce a modification on the cosmic microwave background through the integrated Sachs-Wolf effect. 
Searching for this effect in the Planck data we found a signal of the expected strength, however, with low significance. 
Characterising these superstructures is also important for astrophysical research, for example the study of the environmental dependence of galaxy evolution as well as for precision tests of cosmological models. 
}

   \keywords{Cosmology:observations, Cosmolgy: large-scale strcuture of the Universe, galaxies:clusters:general} 
   \maketitle
%
%-------------------------------------------------------------------

\section{Introduction}

Large efforts have been invested in the study of the cosmic large-scale structure with a main motivation to test cosmological models.
The majority of these studies describe the large-scale structure only in a statistical way (e.g. \citealt{Gro1977,Fry1980}). 
But it is also interesting to map the large-scale structure and to describe the individual properties of the structural elements. 
This was a particularly hot topic when the topology of the cosmic web was discovered (e.g. \citealt{Joe1978,Oor1983,Lap1986}). 
An interesting, modern approach in this field is at present constraint reconstructions of the matter distribution in the local Universe (e.g. \citealt{Hof2015,Jas2019,Dol2023,Lil2024}).

In addition, astrophysical processes, such as galaxy formation and evolution, depend on environmental parameters. 
It is therefore necessary to be able to characterise the large-scale physical environment of a survey field. 
The early discovery of the morphological segregation of galaxy types on supercluster scales by \citet{Gio1986} illustrates this topic. 
Last but not least the density and streaming flow structure of the local Universe has an influence on the measurement of cosmological parameters; for example a local underdensity leads to slightly larger local Hubble constant compared to the cosmic mean (e.g. \citealt{Mar2013,Boe2020}). 
The local matter distribution also affects the cosmic microwave background via the integrated Sachs-Wolfe effect (e.g.~\citealt{Cri1996}). 
Therefore the mapping of the large-scale matter distribution is an important task for observational cosmology.

The universe out to a redshift of $z \sim 0.03$ has been well studied over the entire sky since many years (see e.g. \citealt{Boe2021c} and references therein). 
It is, for example, well covered by the 2MASS redshift survey \citep{Huc2012,Bil2014, Mac2019} and by the Cosmic Flow peculiar velocity compilations (e.g. \citealt{Tul2016,Tul2019,Cou2023}).
In this paper we present an all-sky study of the region between $z = 0.03 - 0.06$ using galaxy clusters as tracers of the large-scale matter distribution. 

Galaxy clusters trace the large-scale matter distribution in a true and slightly amplified, biased, way. 
In order to apply this to map the density distribution of the nearby Universe and to find the largest structures, a comprehensive, statistically highly complete galaxy cluster sample is required with a well-understood selection function. 
Our CLASSIX cluster survey, based on the cluster detection in X-rays in the ROSAT All-Sky Survey \citep{Tru1993}, providing
an almost all-sky coverage, fulfils these requirements. 
It comprises the REFLEX cluster survey in the South \citep{Boe2004,Boe2013} and the NORAS survey in the North \citep{Boe2000,Boe2017}. 
In addition, it includes part of the so-called Zone of Avoidance (ZoA, at galactic latitudes $|b_{II}| < 20^o$) covering in total 86\% of the sky.

We have already demonstrated this method of exploration by assessing the structure of the local Universe at $z \le 0.03$, characterising more well-known superclusters \citep{Boe2020,Boe2021a,Boe2021b,Boe2021c}.
We are now focussing on uncharted territory between $z = 0.03 - 0.06$ where cosmography has not been studied on an all-sky scale before.
In this redshift range the cluster density is still sufficiently high in our survey for an efficient mapping of the matter density. 

The paper is structured as follows. In Sect. 2 we describe the cluster sample.
In Sect. 3 we outline the method of matter density mapping. 
Sect. 4 gives a description of the superstructure construction and the properties of the superstructures.
In Sect. 5 we compare the large-scale structure as traced by galaxy clusters and by galaxies and in Sect. 6 we compare the observational properties of the superstructures to those found in cosmological simulations. 
Sect. 7 contains a discussion of the observational results. The imprint of the superstructures on cosmological observations is outlined in Sect. 8. And Sect. 9 provides conclusions and an outlook. 
We adopt a Hubble parameter of $H_0 = 70$ km s$^{-1}$ Mpc$^{-1}$, a flat cosmology, and a matter density parameter, $\Omega_m = 0.3$.

% ======================================================================
\section{The cluster sample}
% ======================================================================

The {\sf CLASSIX} cluster sample has been established through the X-ray detection of the clusters in the ROSAT All-Sky Survey and their positive identification and redshift measurement through a series of optical spectroscopic follow-up observations in numerous observing campaigns mostly at ESO La Silla, the former German-Spanish observatory Calar Alto, other places including the South African Observatory in Sutherland and through literature data \citep{Guz2009,Cho2012}.
The survey reaches a flux limit of  $1.8 \times 10^{-12}$ erg s$^{-1}$ cm$^{-2}$ in the energy band between 0.1 and 2.4 keV and is highly complete ($> 90\%$) outside the Galactic band ($|b_{II}| \ge 20$ degrees). 
The survey also includes that part of the ZoA where the interstellar hydrogen column density is below $n_H \le  2.5 \times 10^{21}$ cm$^{-2}$ and reaches a conservatively estimated completeness in the cluster identification of $> 70\%$. 
All clusters have spectroscopic redshifts. 
This survey offers the unique advantage that the X-ray luminosity is tightly correlated to the cluster mass \citep{Pra2009,Boe2012}, much better than optical characterisations. 
Our survey covers nearby objects more completely than any other X-ray or Sunyaev-Zel'dovich cluster surveys. 
The eROSITA survey provides a deeper look into the X-ray sky, but so far the public data on galaxy clusters cover only 31\% of the sky with incomplete spectroscopic redshifts \citep{Bul2024}, compared to 86\% sky coverage of {\sf CLASSIX}.

The  cluster sample, that was previously applied to cosmological studies and is used here, was compiled with the above given flux limit and the requirement of at least 20 source photons, for which precise selection functions were published \citep{Boe2013,Boe2017}. 
While in most of the sky the nominal flux limit is reached, there are some regions with low survey exposure and/or high interstellar absorption, where the flux limit is slightly higher.
This is taken into account in the selection function as a function of sky position and redshift and affects about 22\% of the sky with relatively small corrections.

\begin{figure} % Do NOT use \begin{figure*}
	\centering
	\includegraphics[width=0.24\textwidth]{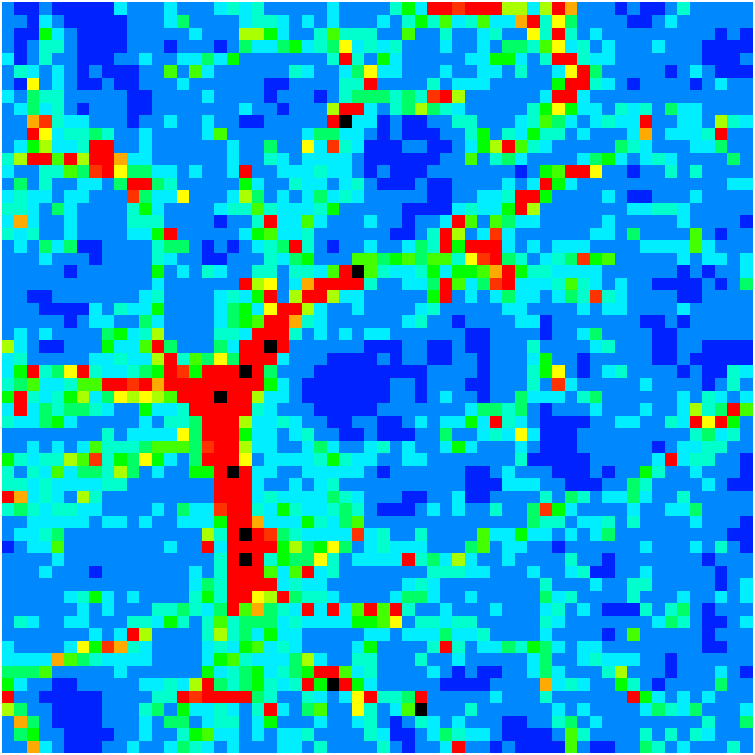}
 	\includegraphics[width=0.24\textwidth]{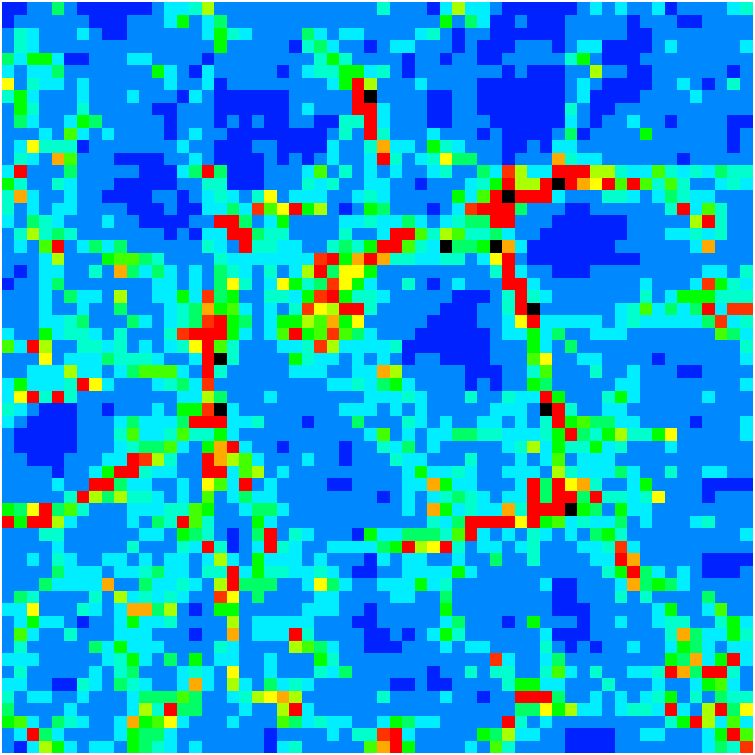}
	\caption{
    Matter density distribution in two regions with a side-length of 117 $h_{100}^{-1}$~Mpc from the Millennium simulation. 
The depth of the slices through the simulation is 15.6 $h_{100}^{-1}$~Mpc. 
The matter density is colour-coded, with increasing density in rainbow colours from blue to red.
The positions of the clusters, selected analogously to the clusters of the extended CLASSIX survey, are marked by black dots.}
\label{fig:mapping} % give each figure a logical label name
\end{figure} 

\begin{figure} % Do NOT use \begin{figure*}
	\centering
    \includegraphics[width=0.44\textwidth]{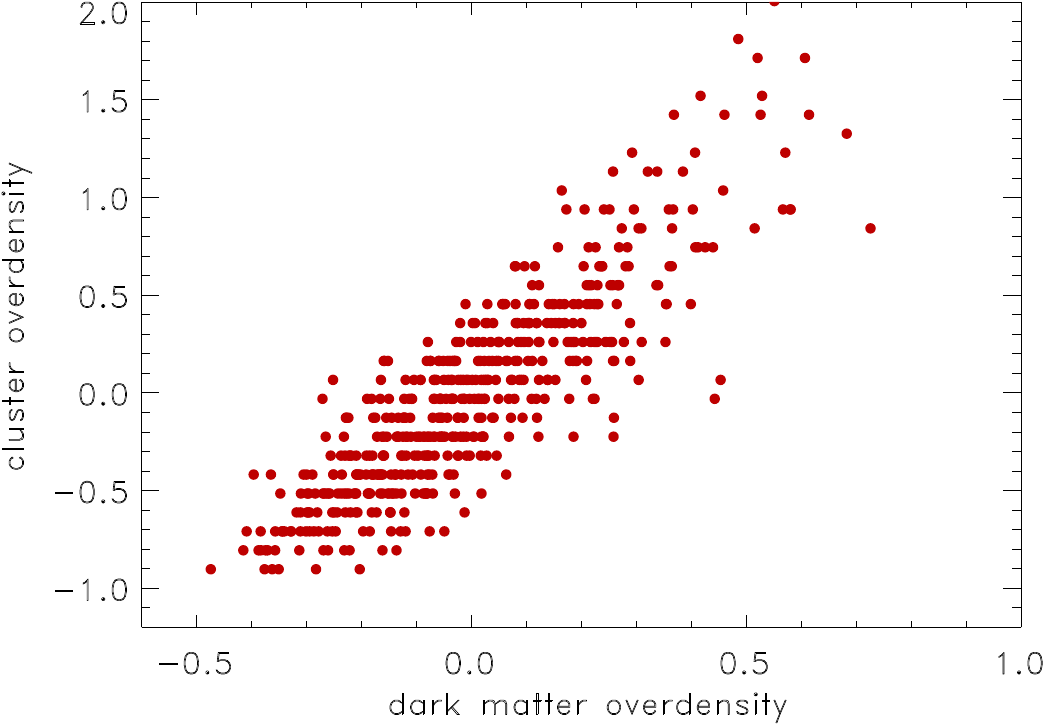}
 	\includegraphics[width=0.44\textwidth]{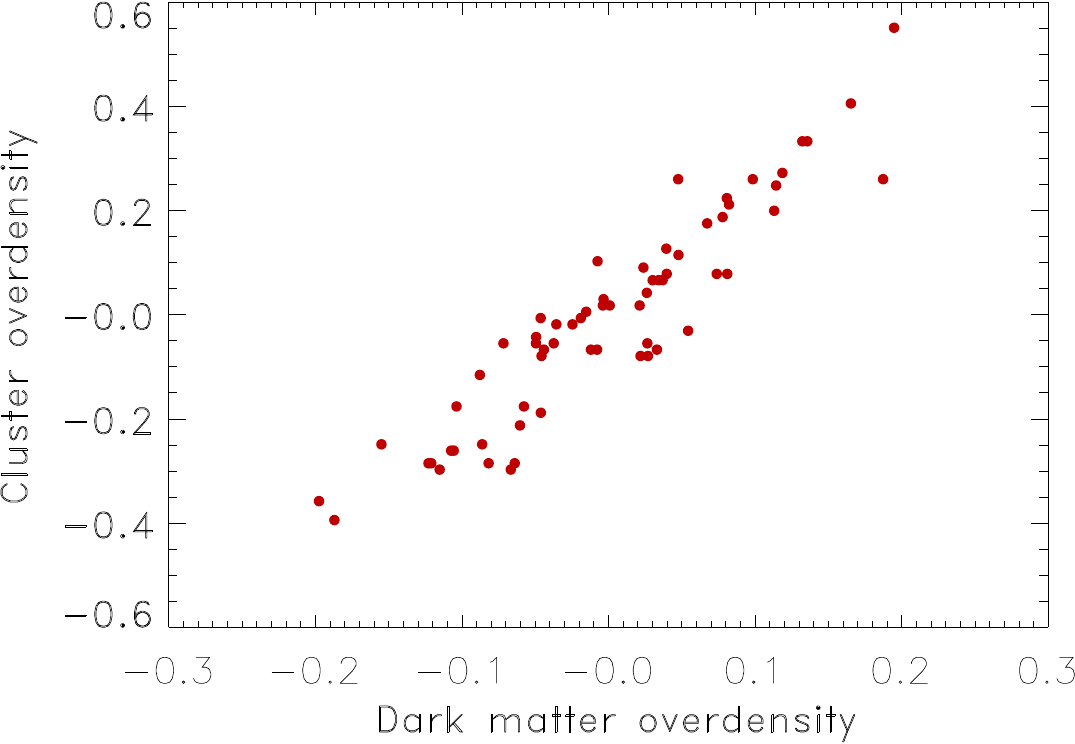}
	\caption{Correlation of the overdensity of matter to that of clusters in cells of 64 $h_{100}^{-1}$~Mpc (top) and $128 h_{100}^{-1}$~Mpc side-length (bottom) in the Millennium simulation.}
\label{fig:Fig2} % give each figure a logical label name
\end{figure} 

In total 345 clusters are contained in the target redshift range, $z = 0.03 - 0.06$, and 155 of these belong to the superstructures described in this paper. 
We also look for clusters which are linked to these superstructures outside the redshift range.
Including these additional clusters the superstructures have 185 cluster members.
Table~\ref{table:cluscat} in the appendix  provides a catalogue of these clusters in superstructures with sky coordinates, redshifts, observed X-ray luminosities, $L_{X,500}$, measured in an aperture of $r_{500}$ \footnote{$r_{500}$ is the radius inside which the mean density of the cluster is 500 times the critical density of the Universe at the cluster redshift.
$r_{200}$ is the analogous radius for a density ratio of 200.}, and mass, $M_{200}$, inside a cluster radius of $r_{200}$,
which has been estimated via the observed X-ray luminosity - mass relation \citep{Boe2013,Pra2009}. 
This mass estimate has a 1$\sigma$ uncertainty of about 40\%.

%% EDITGC - Check the caption again.
\begin{figure} % Do NOT use \begin{figure*}
	\centering
	\includegraphics[width=0.44\textwidth]{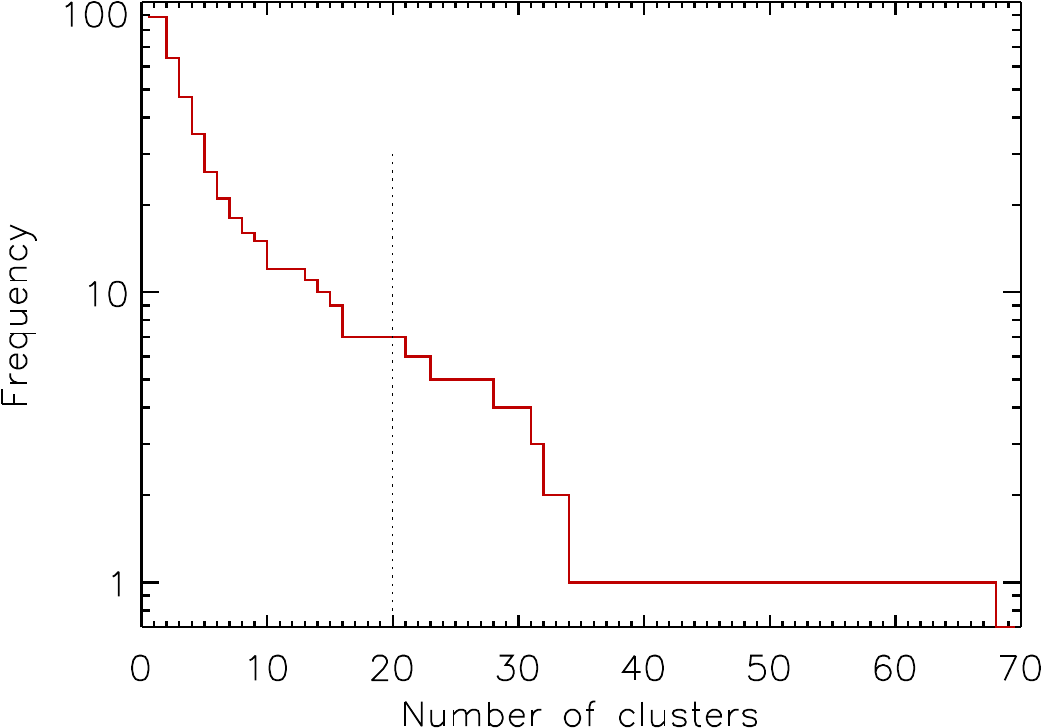}
	\caption{Multiplicity function of clusters in superstructures, that is the distribution of the number of member clusters in these structures. 
    For this accounting we have included also member clusters outside the target redshift interval.
    }
    \label{fig:multi} % give each figure a logical label name
\end{figure}

\begin{figure*}
   \centering
   \includegraphics{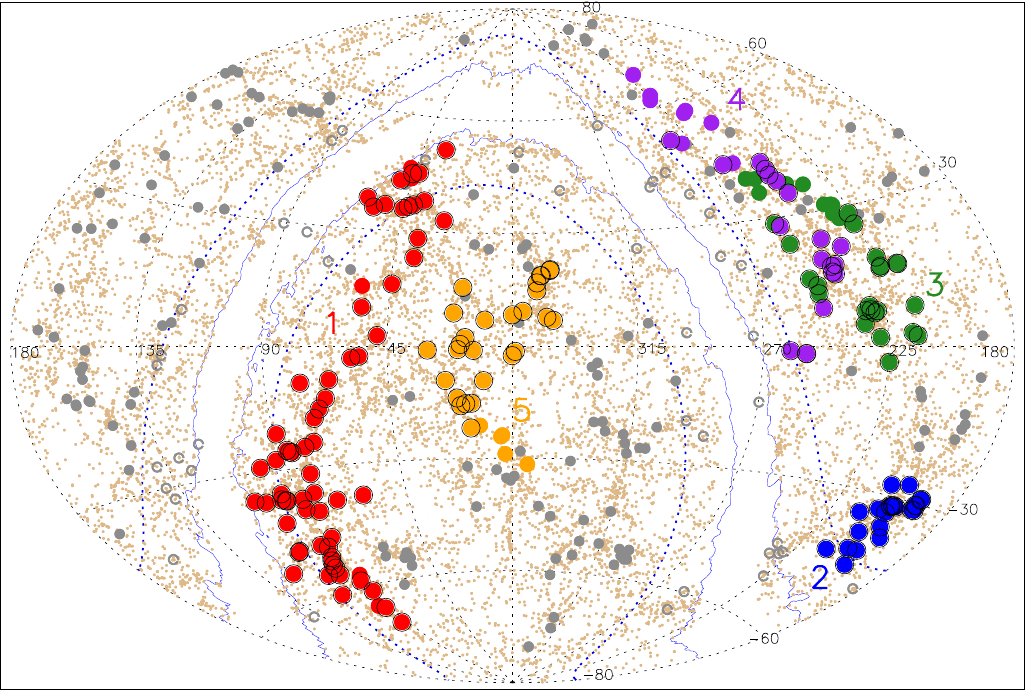}
   \caption{
   Distribution of the CLASSIX clusters in the redshift shell z = 0.03 - 0.06 (solid and open circles).
    The member clusters of the five superstructures are marked in colour, Quipu (red), Shapley (blue), Serpens-Corona Borealis (green), Hercules (purple) and Sculptor-Pegasus (beige). 
    The members of the superstructures outside the target redshift range are not  encircled by a black ring. 
    The clusters outside superclusters are shown as gray circles, with open circles in the region of the Zone of Avoidance.
    The small brown dots show the distribution of 2MASS galaxies in the same redshift region.  
    The dotted blue lines show the Zone of Avoidance region ($|b_{II}| < 20^o$) and the solid blue lines encloses the region with an interstellar hydrogen column density $n_H > 2.5 \times 10^{21}$ cm$^{-2}$.
}
    \label{fig:Fig1}%
\end{figure*}

% ======================================================================
\section{Matter density mapping}
% ======================================================================

We illustrate the connection of the cluster distribution with the underlying matter distribution in Fig.~\ref{fig:mapping}  
by means of the Millennium simulation. 
Shown is the matter density distribution in two slices of the simulation with a thickness of 15.6 $h_{100}^{-1}$~Mpc and an extent of 117 $h_{100}^{-1}$~Mpc. 
The clusters, marked by black dots, all occupy locations in the dense matter filaments.
Fig.~\ref{fig:Fig2} compares directly the cluster and matter density in cubic cells of 64 and 128 $h_{100}^{-1}$~Mpc side-length from the simulations. 
There is a perfect correlation with a scatter that can be well approximated by Poisson uncertainties \citep{Boe2020}.
The proportionality constant of the relation reflects the amplification (biasing) factor of the density contrast.
We have applied this correlation already successfully for the large-scale structure mapping in the local Universe \citep{Boe2020,Boe2021a,Boe2021b,Boe2021c}, where we found, for example, a local underdensity in the matter distribution \citep{Boe2020}, that is also confirmed by other methods \citep{Whi2014,Jas2019,Dol2023}.

% ======================================================================
\section{Constructing and characterising the superstructures}
% ======================================================================

To unveil the largest matter concentrations, we searched for superstructures in the local Universe comprising at least 20 {\sf CLASSIX} cluster members and with overdensities, $\Delta = (\rho - \bar\rho)/\bar\rho)$, in the cluster distribution of the order of one, which clearly evolved away from the linear regime of small density amplitudes.
These superstructures have larger sizes and lower overdensities than the typical superclusters and are comparable only to the largest superclusters known \citep{Cho2015}. 
To find the superstructures we applied a friends-of-friends algorithm.

Because the cluster sample was constructed with a flux-limit, the cluster density changes with redshift. 
To select similar superstructures at different redshifts, the linking length was adjusted to the spatial density of the clusters. 
With the aim to obtain structures with an overdensity in the cluster distribution of at least $\Delta = 1$, we adjusted the linking length, $l$, to the cluster density, $N_{Cl}$, as $l = \left( 2 \cdot N_{Cl} \right) ^{-1/3}$. 
The cluster density was locally estimated by means of the cluster X-ray luminosity function and the local parameters of the selection function.
Due to the adjustment to the redshift dependent cluster density, the linking length covers the moderate range of $l = 26 - 56$~Mpc, while the mean value of the linking length is 38.5~Mpc.

As we are interested only in the largest structures we chose a threshold of minimum 20 members for the selection. An inspection of the multiplicity function shown in Fig.~\ref{fig:multi} can provide a further motivation for this choice. 
This function shows the distribution of the number of member clusters for all superclusters down to pairs.
We note that the next smaller structure containing at least half their members in the target volume has 14 cluster members. 
Two further superstructures with 15 and 16 members have 7 member clusters in the target volume. 
Thus the cut at 20 members falls on a plateau in the multiplicity function and we would have only found more superstructures with a considerably lower cut. 

Allowing the linking to reach beyond the above defined redshift range (to allow for a complete recovery of each structure) we found five superstructures with our requirements. 
At lower redshift only the well-known Perseus-Pisces Supercluster \citep{Joe1978} fulfils the search criteria. 
We have already described this structure in detail in a previous publication \citep{Boe2021b}.

\begin{figure*}
   \centering
   \includegraphics[width=0.98\columnwidth]{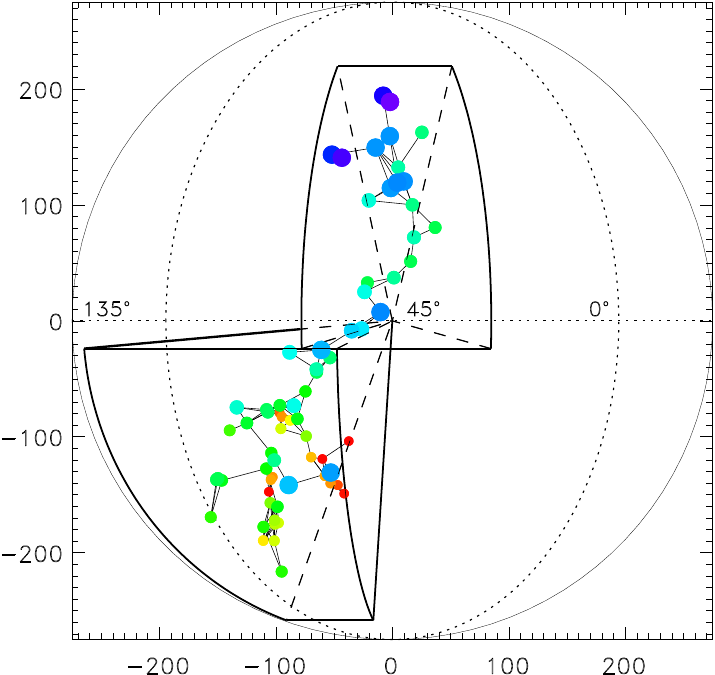}
   \includegraphics[width=0.98\columnwidth]{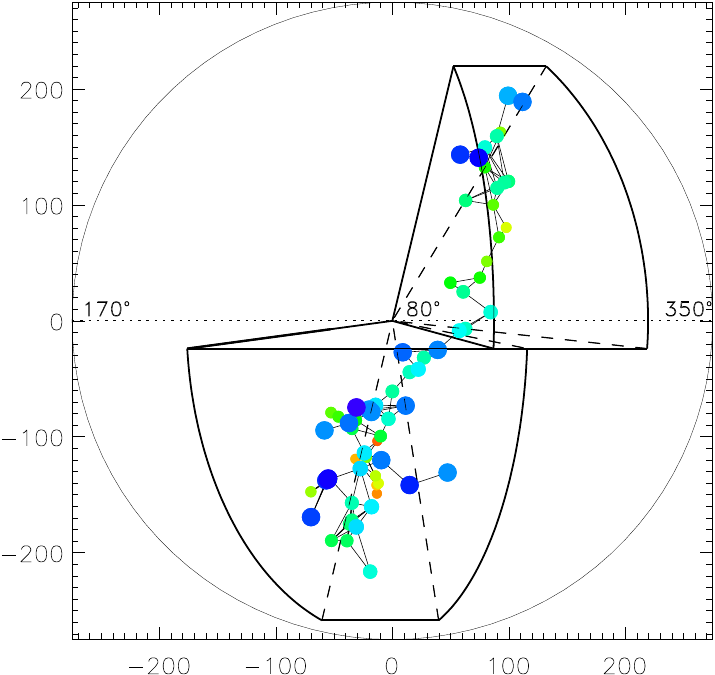}   
   \caption{
   Three-dimensional representation of the Quipu Superstructure in equatorial coordinates.
   The wedges indicate the embedding volume of the structure separately for the northern and southern sky. 
   The dots are the cluster members with sizes and colours indicating the distance to an external observer, blue for the nearest and red for the most distant clusters. 
   The connecting lines show the percolation links.
   We present two views of the structure, in the left panel we face the northern wedge with (at RA = 45$^o$ and in the right panel the southern one (at RA = 80$^o$ in the centre).}
    \label{fig:Quipu}
\end{figure*}

\begin{figure}
   \centering
   \includegraphics[width=0.74\columnwidth]{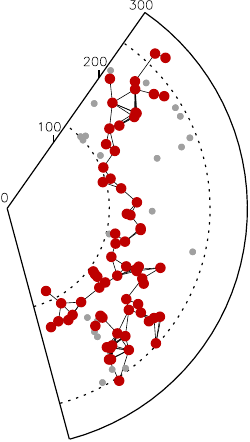}
   \caption{
   Wedge diagram in declination and distance of the Quipu superstructure. The distance is in units of Mpc. 
   The red dots show the superstructure members and the black lines show the friends-to-friends linking. 
   The grey dots show the non-member clusters in the target redshift range and right ascensions between 27.7$^o$ and 119.4$^o$. 
   The two dashed lines give the distances for redshifts of 0.03 and 0.06.}
    \label{fig:Quipu2}
\end{figure}

Table\,1 provides information on the properties of the five superstructures.
Their estimated mass ranges from  0.6 to $2.4 \times 10^{17}$ M$_{\odot}$.
These masses were estimated by assigning a volume to the superstructures multiplied by the implied matter overdensity. 
To determine the latter we used the observed cluster overdensity and applied a bias factor between the matter and cluster distribution of about 1/3 - only for Hercules, which has a significantly lower mean mass of the clusters, we use a factor of 1/2. 
These bias factors are obtained from theory \citep{Tin2010} and from observations of the {\sf CLASSIX} sample \citep{Bal2011,Cho2014}.  
The volume assignment is less straightforward. 
A plausible solution is to take the sum of all volume elements with a distance between half and full linking length to any member cluster. 
We obtain matter overdensities of $\Delta \sim 1$ with a distance limit of 0.75 of the mean linking length (28.875 Mpc) and adopted this assignment.
The length of the structures is determined from the largest distance between any of the member clusters.    
%..............................................................
\begin{table*} 
	\caption{Properties of the superstructures found in the present study.}
	\label{tab:Tab1} 
	\begin{tabular*}{\textwidth}{@{}llccccllll} 
\hline
name & $N_{Cl}$& $<z>$ & z-range & RA     & DEC & length & $M_{est}$ & $\Delta _{Cl}$ & $\Delta _{DM}$ \\ 
\hline
1. Quipu      & 68 (63) & 0.043 & 0.027 - 0.065 & 27.7 - 119.4  & -69.8 -- 51.7  &  428 & 2.4  & 1.4 & 0.5\\
2. Shapley    & 23 (23) & 0.049 & 0.038 - 0.056 & 193.1 - 213.1 & -44.1 -- -27.2 & 90  & 0.8  & 3.8 & 1.3 \\
3. Ser-CorBor & 34 (23) & 0.056 & 0.035 - 0.078 & 214.4 - 257.8 & -3.5 -- 39.7  & 234 & 1.8  & 2.7 & 0.9   \\
4. Hercules   & 28 (19) & 0.032 & 0.024 - 0.046 & 234.9 - 279.0 & -1.8 -- 68.1  & 154 & 0.6  & 0.7 & 0.3\\
5. Scu-Peg    & 32 (27) & 0.047 & 0.037 - 0.065 & 345.3 - 30.6  & -31.6 -- 20.3 & 216 & 1.3  & 2.1 & 0.7\\
\hline
%\hline
\end{tabular*}
\smallskip

{\bf Notes:} $N_{Cl}$ is the number of cluster members, the numbers in brackets give the number of members inside $z = 0.03 - 0.06$, $<z>$ is the mean redshift, the length is in units of Mpc, $M_{est}$ is the estimated supercluster mass in units of $10^{17}$ M$_{\odot}$, $\Delta _{Cl}$ and $\Delta _{DM}$ are the cluster and matter overdensity.
The numbering that appears with the name of the superstructures is used for the labelling of the figures throughout the paper.
\end{table*}

\subsection{The Quipu superstructure}

Figure~\ref{fig:Fig1} shows the location and shape of the superstructures in the sky. 
The largest and most impressive object is the Quipu superstructure \footnote{Named after the bundles of cords with knots used by the Inkas
to store administrative and calendrical information, which resemble this superstructure in their shape. 
The name was also chosen, since most the cluster redshifts have been obtained at ESO in Chile where such records were found.}.
With a length of 428~Mpc and an estimated mass of $\sim 2.4 \times 10^{17}$ M$_{\odot}$ it is significantly larger than any previously known structure.  
In the north, Quipu ends at the obscured region in the ZoA and it is possible that the superstructure extends further into the ZoA, in particular since we note a concentration of clusters at the other side of the ZoA. 
In the south, the edge of Quipu is located very close to the Vela supercluster in the ZoA. 
The two structures would have been linked with a 16\% larger linking length.
The next subsection provides more details about the possible connection of Quipu to the Vela supercluster.
Figure~\ref{fig:Quipu} shows Quipu in two different three-dimensional representations. In these displays the clusters are projected on an x and y plane and we look at them from the z-direction which points to right ascension 45$^o$ in the left panel and 80$^o$ in the right right panel. 
The closer the clusters are to the observer in the z-direction the larger their size, while the colours change with increasing distance in rainbow colours from blue to red. 
We perceive the superstructure as a long filament with side extensions. 
The 20 clusters in the northern hemisphere have a mean right ascension of 45.5$^o$.
In the south the filament bends to the east and the 48 southern clusters have a mean right ascension of 89.3$^o$. Thus in the left panel of Fig.~\ref{fig:Quipu} we are facing the northern part of Quipu and in the right panel the southern part in the centre.

In Fig.~\ref{fig:Quipu2} we show this superstructure again as a wedge plot in
declination and redshift direction. This view gives the best impression of the superstructure as a long filament with small side filaments which initiated the naming of Quipu.
Quipu is actually a prominent structure readily noticeable by eye in a sky map of clusters in the target redshift range, without the help of a detection method.
It can also easily be recognised in the map of the galaxy distribution shown in the next section.

\subsection{The other superstructures}

\begin{figure} 
	\centering
	\includegraphics[width=0.31\textwidth]{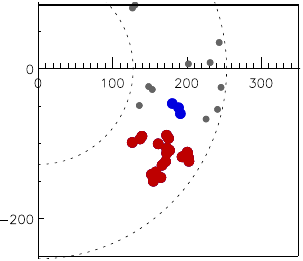}
 	\includegraphics[width=0.30\textwidth]{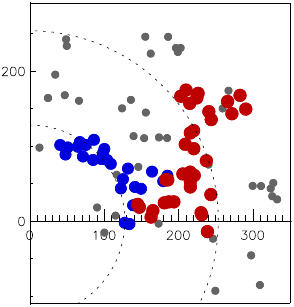}
	\includegraphics[width=0.30\textwidth]{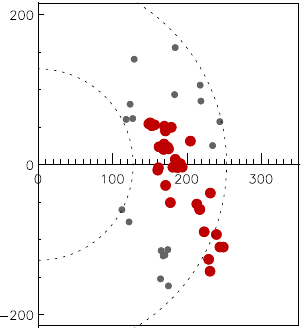}    
	\caption{
    Four of the superstructures shown in declination-redshift diagrams. 
    The distance units are in Mpc and the two dotted lines show the redshift distance  $z = 0.03$ and $0.06$. \\
    {\bf Top:} Shapley supercluster shown by red dots. 
    Grey points show the non-member clusters in the same redshift range with right ascensions between 190$^o$ and 215$^o$. 
    The clusters marked by the three blue dots are linked to the Shapley supercluster for an increased linking length (see text).\\   
    {\bf Middle:} Serpens - Corona Borealis, marked in red and Hercules, marked in blue.
    Grey points show the non-member clusters in the same redshift range with right ascensions between 214.4$^o$ and 279$^o$.  \\    
    {\bf Bottom:} Sculptor-Pegasus superstructure shown by red dots.  
    Grey points show the non-member clusters in the same redshift range with right ascensions between 345.3$^o$ and 360$^o$ as well as 0$^o$ and 30.6$^o$.
 }
\label{fig:Shapley} % give each figure a logical label name
\end{figure}

Another remarkable structure is the well-known, massive and compact Shapley supercluster. 
It has the smallest size in our sample and is the most massive supercluster identified
to date \citep{Rei2000,Cho2015}. 
It shows a remarkable isolation, indicated by the fact that no more than three members are added if the linking length is increased by up to 25\% and it is stable against a decrease 
of the linking length by $ \le 30\%$. 
The Shapley supercluster is shown in form of a wedge diagram with redshift and declination axes in the top panel of Fig.~\ref{fig:Shapley}. 
The three clusters which are associated to the supercluster with an increase of the linking length by a factor of 1.08 are shown as blue dots. 
The empty region around Shapley is well illustrated by this figure. 
In Table~\ref{tab:Tab1} the Shapley supercluster features the highest overdensity. 
This high overdensity and isolation of Shapley indicates that a lot of matter from the surroundings has been swept up to form this massive and dense supercluster.

\begin{figure} 
	\centering
	\includegraphics[width=1.0\columnwidth]{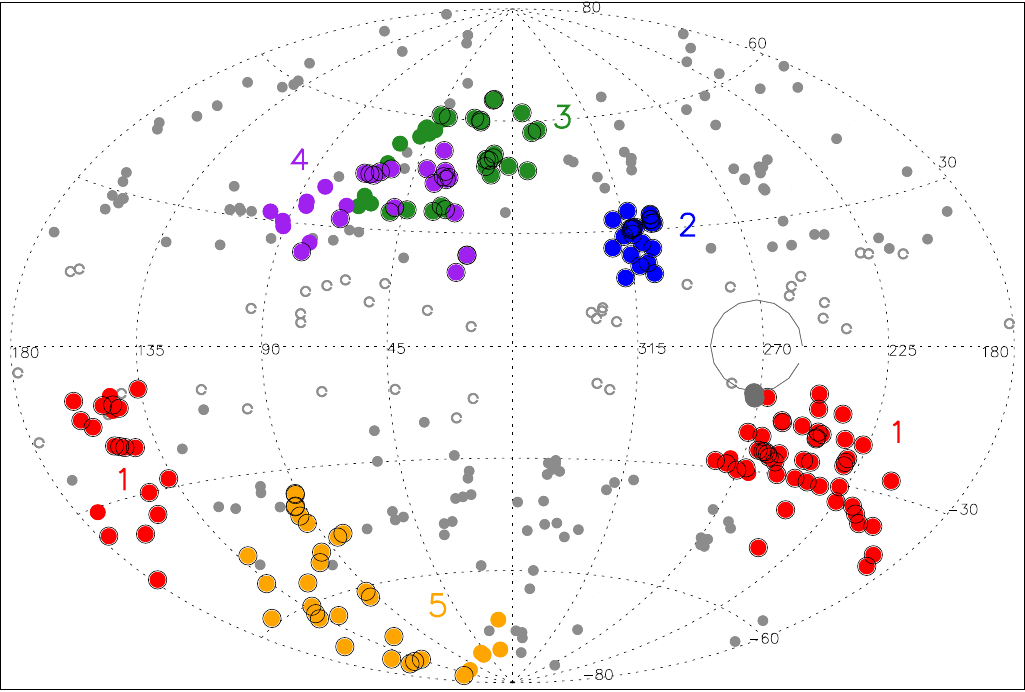}
	\caption{
    Distribution of the five superstructures on the sky in galactic coordinates, with the same colours and symbols as in Fig.~\ref{fig:Fig1}. 
    In addition, the grey, open circle shows the location of the Vela supercluster in the ZoA. 
    Two grey, filled circles (overlapping in the figure) mark the two X-ray luminous clusters identified in the ROSAT All-Sky Survey at the edge of this supercluster.
 }
\label{fig:skymap2} % give each figure a logical label name
\end{figure} 

Serpens - Corona Borealis and the extended Hercules superclusters are overlapping in the sky but they are well enough separated in redshift and are not percolated. 
This can be seen in the wedge diagram in the middle panel of Fig.~\ref {fig:Shapley}. 
Hercules is larger in our survey than described previously by \citet{Bar1998}. 
It is made up of many low mass systems and has the lowest overdensity.
The Hercules supercluster has a large extension into the redshift region $z \le 0.03$, while Serpens - Corona Borealis extends to higher redshifts.

The fifth superstructure extends from the constellation of Sculptor to Pegasus from which it gets its name. 
Five of the member clusters are located at redshifts $z > 0.06$. 
The Sculptor - Pegasus superstructure, which is shown in a redshift declination diagram in the bottom panel of Fig.~\ref{fig:Shapley}, appears as an elongated filament.

% ======================================================================
\subsection{The connection of Quipu to the Vela supercluster}
% ======================================================================

As noted above, the Quipu superstructure comes close to the Vela supercluster, which is buried in the ZoA. 
This supercluster was identified at the location $l = 272.5^o \pm 20^o, b= 0^o \pm 10^o$ and $z \sim 0.06$ in the galaxy distribution of a redshift survey based on the 2MASS survey and dedicated deep optical surveys \citep{Kra2017}. 
It was also identified as a dynamically important mass concentration in an analysis of the cosmic flow traced by galaxy proper motions from redshifts and Tully-Fisher distances \citep{Cou2019, Mou2024}.
In X-rays two galaxy clusters can be identified in the outskirts of this supercluster RXCJ0812.4-5714 (CIZA0812.5-5714) at $z = 0.0619$ and RXCJ0820.9-5704 (CIZA0820.9-5704) at $z = 0.0610$, which are included in the extended CLASSIX catalogue used for the construction of our superstructures. 
With a linking length about 16\% larger than what was used in our study, the two Vela supercluster members would have been part of Quipu.

Fig.~\ref{fig:skymap2} shows the location of Vela and the two member clusters with respect to Quipu in a galactic coordinate representation of the sky.  
Comprehensive ongoing galaxy surveys in neutral hydrogen in the ZoA trace the galaxy wall of the Vela supercluster and another wall in this region at a redshift of $\sim 0.04$, e.g. \citep{Raj2024a,Raj2024b}.
In summary, the Vela supercluster is close, but not linked in our analysis to Quipu, but there is the possibility that new X-ray luminous member clusters could be detected in the ZoA leading to a linking. 
This illustrates that we already have indications about prominent structures in the hidden region of the ZoA, where ongoing surveys will soon provide us with more details.

% ======================================================================
\section{Comparison to structures in the galaxy distribution}
% ======================================================================

Galaxies are also used as tracers of the large scale structure. 
In this section we explore if the galaxy distribution shows the same superstructures that we find with clusters of galaxies. 
Currently the best all-sky galaxy redshift survey for this purpose is the 2MASS redshift survey \citep{Huc2012,Mac2019}. 
The galaxies are selected by their near infrared brightness minimising the bias in relating stellar mass to luminosity. 
The galaxy sample is highly complete to magnitude limit $K_s \le 11.75$. Fig.~\ref{fig:Fig1}, which shows the {\sf CLASSIX} clusters and 2MASS galaxy distribution, gives a first impression that both tracers show a very similar large-scale distribution.
We note that all five superstructures are well marked by high densities in the galaxy distribution.

\begin{figure}
   \centering
   \includegraphics[width=0.98\columnwidth]{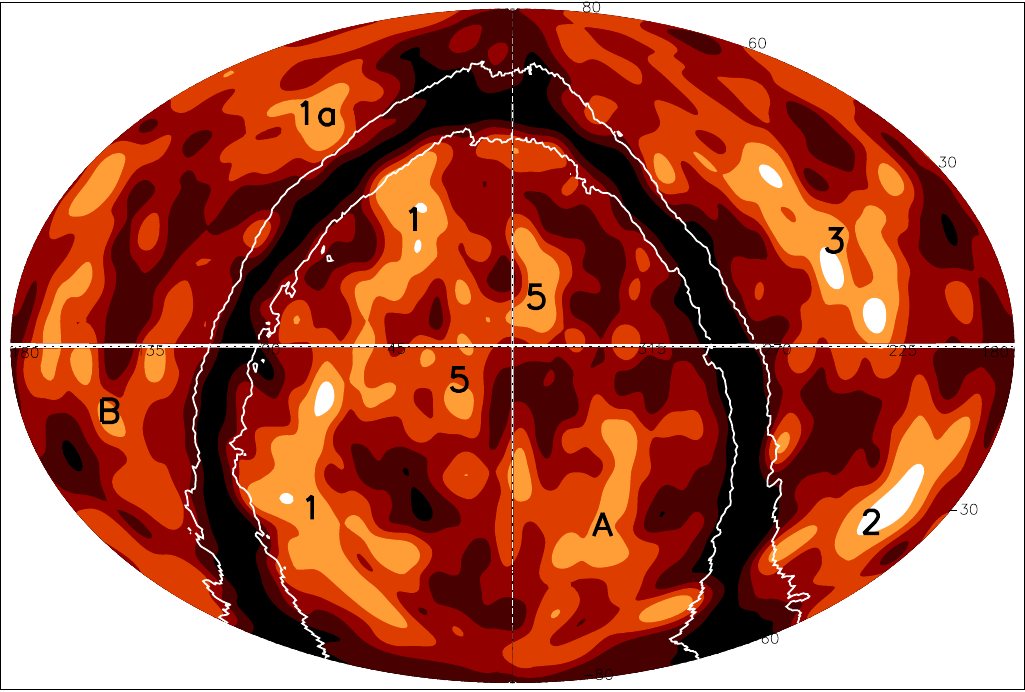}
   \caption{
   Density map of the 2MASS galaxy distribution in the redshift interval $z = 0.03 - 0.06$ in equatorial coordinates. 
   The density ratio to the average density is shown by six contour levels: 0 - 0.23 (black), 0.23 - 0.62 (dark brown), 0.62 - 1.13 (middle brown), 1.13 - 1.9 (light brown), 1.9 - 3.7 (orange), and $>$ 3.7 (white). 
   The distribution was smoothed with a Gaussian with $\sigma = 3$ degrees. 
   The five superstructures are labelled with the number as they appear in Table~\ref{tab:Tab1}, except that the overlapping structures of 3 and 4 just carry the label 3. The extension of Quipu on the other side of the Zone of Avoidance is labeled 1a.
   Two more regions showing an enhanced galaxy density are labelled A and B and their identification is discussed in the text. 
   The region around the galactic plane appears black since there are no data on the galaxy distribution in the 2MASS catalogue. 
   The two white lines enclose the region of the ZoA, where the interstellar hydrogen column density is larger than $N_H \ge 2.5 \times 10^{21}$ cm$^{-2}$.}
\label{fig:2Mgal}
\end{figure}

To provide a view on the galaxy distribution without the bias of the superstructures of galaxy clusters, we show in Fig.~\ref{fig:2Mgal} only the 2MASS galaxy distribution in the form of a surface density map smoothed with a Gaussian with $\sigma = 3$ degree. 
We note three prominent, large concentrations, which can easily be identified with Quipu, Shapley and the two superstructures seen in projection, Serpens - Corona Borealis and Hercules, labelled as 1, 2, and 3, respectively.
The possible extension of Quipu at the other side of the ZoA (labeled 1a) is very clearly observed in the galaxy distribution.
The Sculptor-Pegasus superstructure, which is marked by the numeric 5 in the Figure, is a bit less pronounced. 
Two further, similar galaxy density enhancements are labelled as $A$ and $B$ in the map. 
We can identify these structures with the superstructures which failed the criterion to contain more than 20 cluster members, as explained below.

\begin{figure}
 \includegraphics[width=0.49\textwidth]{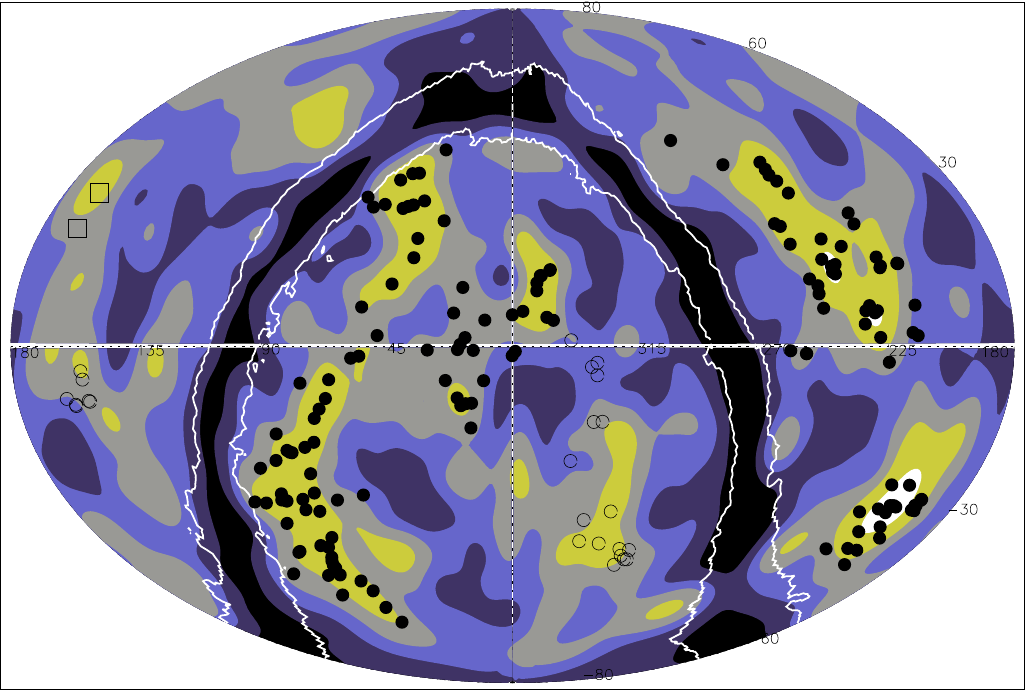}
    \caption{
    Map of the 2MASS galaxy distribution smoothed with a Gaussian of $\sigma = 4$ degrees.
    The density ratio to the average density is shown by six contour levels: 0 - 0.23 (black), 0.23 - 0.62 (dark blue), 0.62 - 1.13 (light blue), 1.13 - 1.9 (grey), 1.9 - 3.7 (olive), and $>$ 3.7 (white). 
    The clusters of the five superstructures are overplotted with filled black circles. 
    The member clusters of two superclusters in region A and one supercluster in region B (as labelled in Fig.~\ref{fig:2Mgal}) are shown with open circles. 
    Two clusters extending into the target redshift range from the Coma supercluster at lower redshift are marked by open squares.}
    \label{fig:gald} 
\end{figure}

In Fig.~\ref{fig:gald} we show the galaxy distribution again, now smoothed with a Gaussian of $\sigma = 4$ degrees and a different colour table that emphasises more dense regions.
We also show the member clusters of the five superstructures as filled black circles. 
The clusters follow the galaxy overdensities impressively well. 
We also note that the galaxy density distribution falls off quite slowly from the region traced by the clusters. 
This is especially interesting in the case of the Shapley supercluster. 
The very compact concentration of clusters is surrounded by a much larger halo of enhanced galaxy density. 

\begin{figure}
   \centering
   \includegraphics[width=0.98\columnwidth]{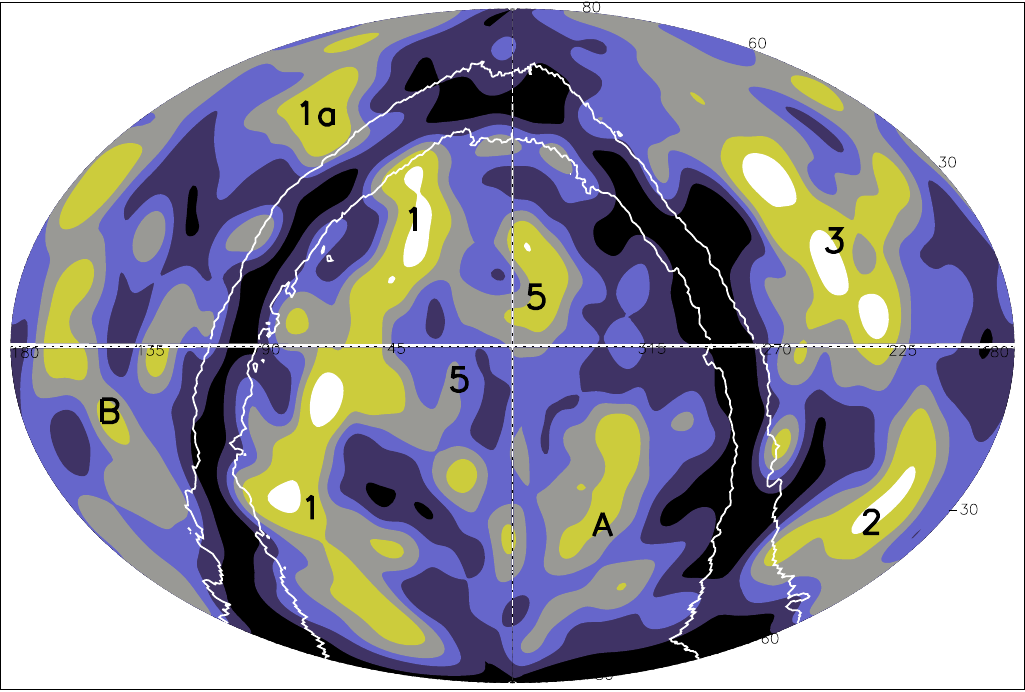}
   \includegraphics[width=0.98\columnwidth]{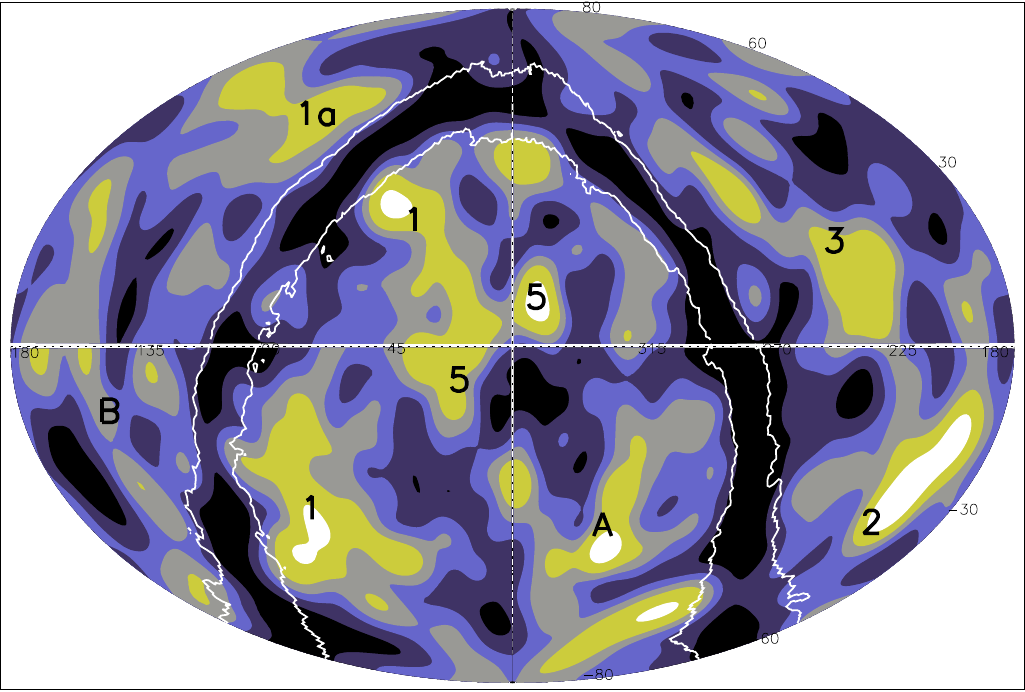}
   \includegraphics[width=0.98\columnwidth]{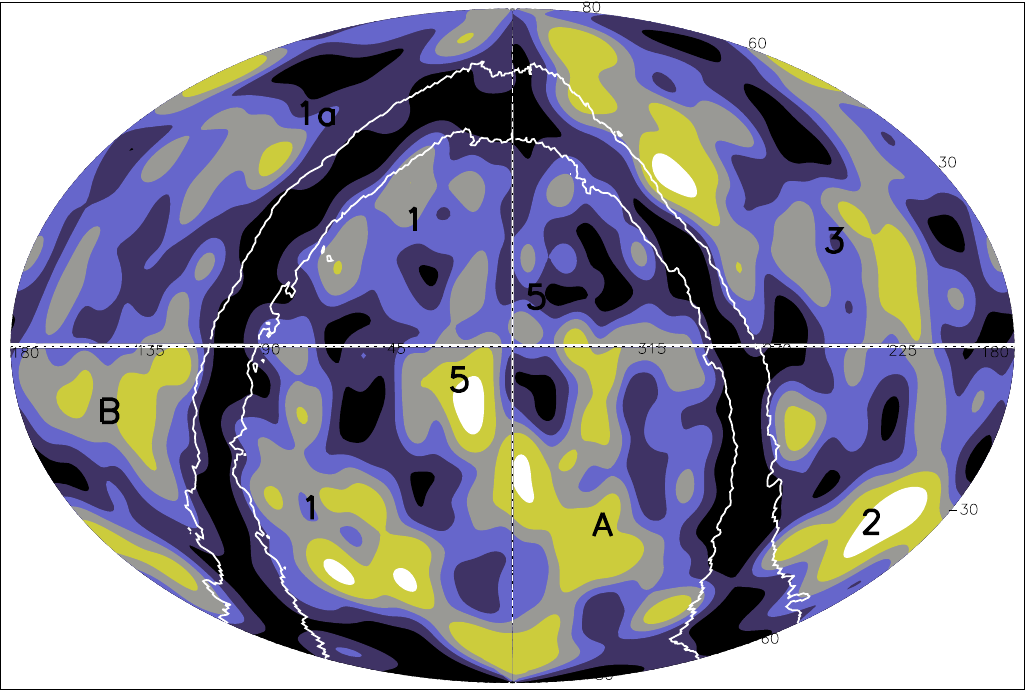}
   \caption{
   Density map of the 2MASS galaxy distribution in the redshift intervals $z = 0.03 - 0.04$ (top), $z = 0.04 - 0.05$ (middle), and $z = 0.05 - 0.06$. The contour levels for the density ratio to the average density are the same as in Fig.~\ref{fig:gald}. 
   The labels and the lines are identical to those in Fig~\ref{fig:2Mgal}.}
    \label{fig:2Mgal_3x}
\end{figure}

We also show the clusters at $z= 0.03 - 0.06$, which belong to the three superstructures with less than 20 members, mentioned above, with open circles in Fig.~\ref{fig:gald}.
In the southern part of region A we show the nine members in the target redshift range of a superstructure with 14 members, and in the northern part we show seven clusters at $z = 0.03$ to $0.06$ of a superstructure with 16 members.
While the southern clusters coincide with the galaxy density peak, the clusters of the northern superstructure are only loosely associated to the galaxy density enhancement. 
In the centre of region B, we see the seven clusters in the target redshift range of the third superstructure with 15 members. 
Thus we can actually associate all large-scale galaxy overdensities with large structures we found in our superstructure construction. 
An exception is a small density patch on the edge of the upper left quadrant, which is marked by two open squares. 
This structure can be identified with two clusters belonging to the Coma supercluster. 
This is just the high redshift tail of the Coma supercluster, sometimes also called Great Wall with most of its mass at $z \le 0.03$.

To show more details of the superstructures in the redshift direction, we display in Fig.~\ref{fig:2Mgal_3x} maps of the 2MASS galaxy distribution for three redshift shells at $z= 0.03 - 0.04$, $z = 0.04 - 0.05$, and $z=0.05 - 0.06$.  
We note that Quipu is well connected only in the lowest redshift shell. 
At intermediate redshifts we still see the northern part with the extension on the other side of the ZoA and the southern part.
This southern part is then also present at the highest redshifts. 
Shapley extends over the entire redshift range, which is not surprising, looking at the redshift distribution of the Shapley cluster members in the top panel of
Fig.~\ref{fig:Shapley}. 
Sculptor-Pegasus stretches as a filament from low redshift in the north to high redshift in the south, as also seen in the bottom panel of Fig.~\ref{fig:Shapley}.

\begin{figure}
   \centering
   \includegraphics[width=0.98\columnwidth]{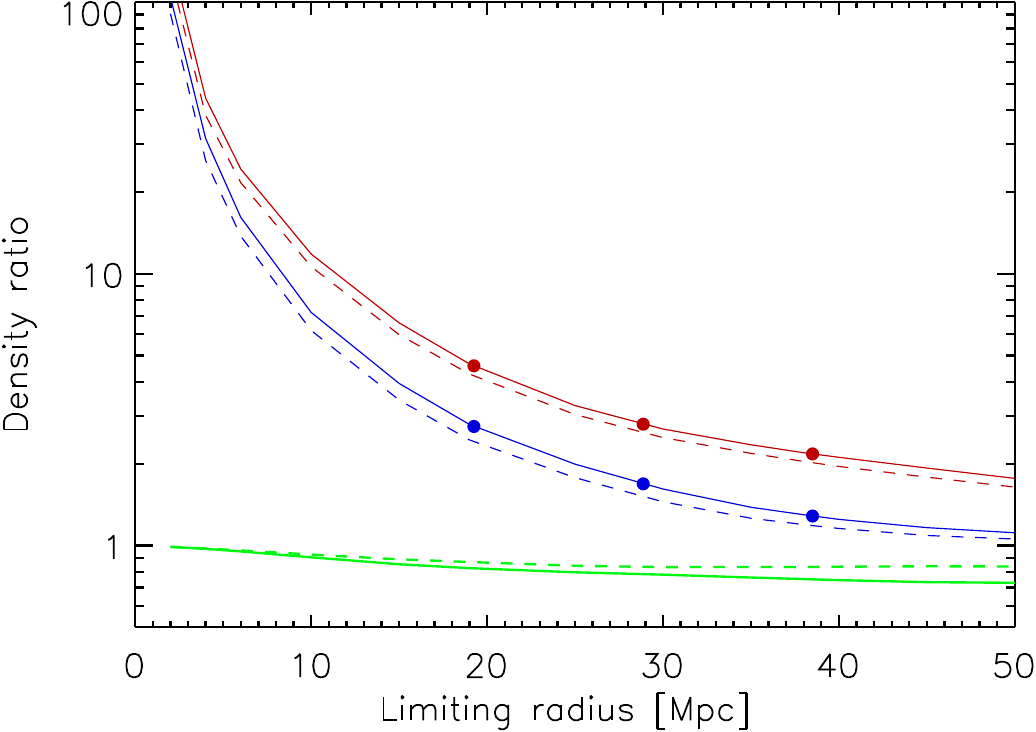}
   \caption{Density ratio to the mean density of the galaxy density distribution in superstructures (red) and in regions around clusters outside superstructures (blue) as a function of the limiting radius. 
   The solid lines show the results for the total survey volume and the dashed lines the ratio for this volume without the Zone of Avoidance.
   The radii for which the values are listed in Table~\ref{tab:densities} are marked with solid points.
   The green lines show the density ratio to the mean density of the regions outside the superstructures (solid line) and outside the cluster halos of the non-member clusters (dashed lines).}
    \label{fig:densprof}
\end{figure}

\begin{figure}
   \centering
   \includegraphics[width=0.98\columnwidth]{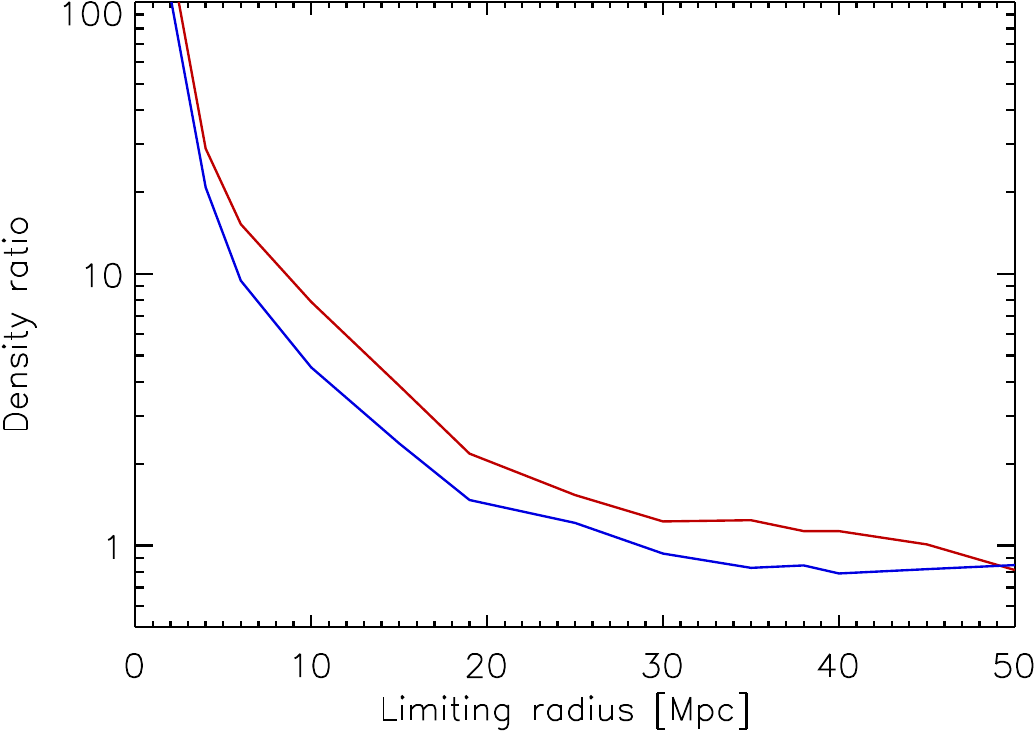}
   \caption{Differential density ratio to the mean density of the galaxy distribution in superstructures (red) and in regions around clusters outside superstructures (blue) as a function of the limiting radius.}
    \label{fig:diffprof}
\end{figure}

To quantify the correlation of the cluster and galaxy distribution, we map the galaxy density as a function of the radial distance from the superstructures in the following way. 
All 2MASS galaxies at $z = 0.03$ to $0.06$ are counted as part of the superstructure, whose distance to the nearest member cluster is smaller than a certain limit. 
We will call this distance in the following the limiting radius. 
We then calculate the mean galaxy density inside the volume determined by the limiting radius. 
The resulting mean galaxy densities as a function of the limiting radius are shown in Fig.~\ref{fig:densprof}. 
We only consider the target volume of the study. 
Since the cluster sample is not so complete in the volume of the ZoA, we also calculated the density distribution function excluding this volume. 
Both results are very similar, with the ZoA-excluded region showing slightly smaller densities. 
The density function falls steeply with increasing limiting radius for smaller radius values followed by a long tail.

To see if the distribution in superstructures is special, we repeat this exercise for all the clusters outside superstructures in the target volume. 
Also here we perform the calculation for the entire target volume and for the region excluding the ZoA. 
These density distribution functions are shown as blue lines in Fig.~\ref{fig:densprof}. 
Both radial density distribution function, those for clusters in superstructures and those outside, start at similar density values in the cluster outskirts, but then they show a significant difference on large-scale, with superstructures living in a significantly more extended overdensity region.

Figure~\ref{fig:densprof} also shows the galaxy density in the entire region that is not occupied by superstructures as a function of the volume attributed to the superstructures as given by the limiting radius. 
A similar exclusion function for all the non-member clusters is also shown a dashed green line. 
We see that the density outside the superstructures decreases with the limiting radius.
The larger we make the volume of the superstructures that is excluded, the smaller is the density in the remaining zone. 
This illustrates again the large extent of the galaxy overdensity region around superstructures. 
The decrease is less pronounced for the clusters unassociated with superstructures as shown by the dashed green line.

In Fig.~\ref{fig:diffprof} we show the galaxy density distribution in differential form. 
We note that we find an overdensity in the galaxy distribution around superstructures up to a limiting radius of about 45~Mpc, whereas the region outside becomes underdense. 
For clusters unassociated with superstructures, the overdensity region extends only to about 30 Mpc. 
Thus we find in all this diagnostics that superstructures define a region that differs from a region that is just a loose collection of clusters.
The results from this section are also summarised in Table~\ref{tab:densities} for three selected radii.

% ======================================================================
\section{Comparison to simulations}
% ======================================================================

\begin{figure}
   \centering
   \includegraphics[width=0.98\columnwidth]{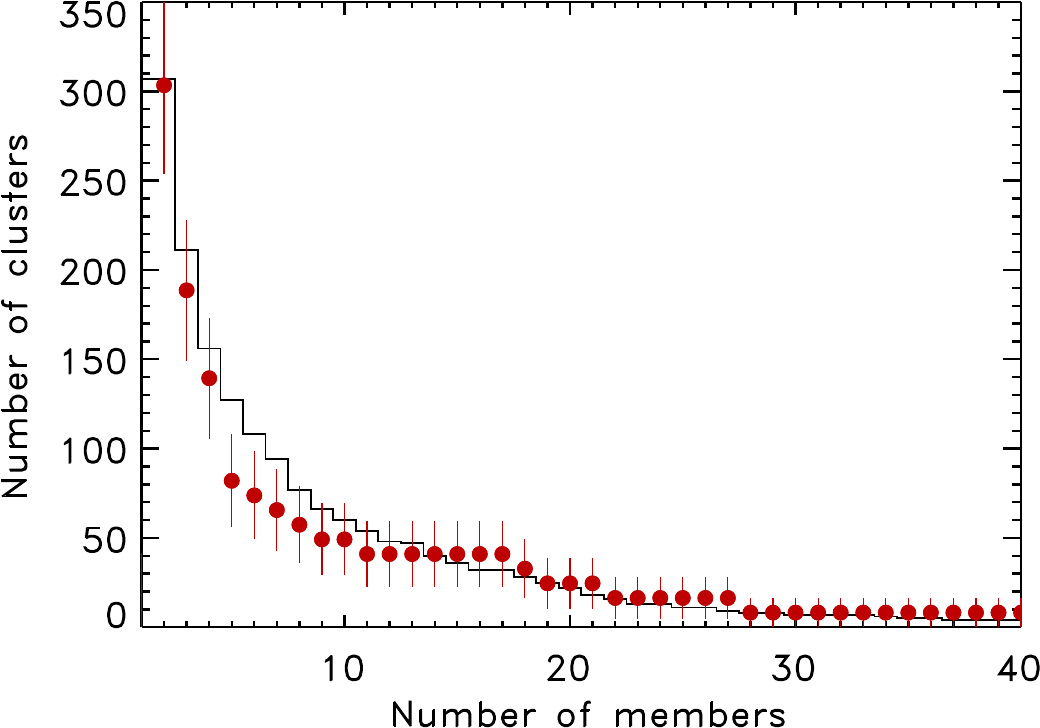}
   \caption{Cumulative multiplicity function of the number of members in observed superstructures (data points) and those from the simulation (black line). The observed distribution has been normalised by the ratio of the parent volumes of the two samples.}
              \label{fig:comp_multi}
\end{figure}

For a better understanding of our findings, we studied analogous objects in the cosmological Millennium simulation \citep{Spr2005}. 
Verifying that the cluster mass function in the simulation is similar to the observed one, we selected clusters from the simulations with a mass limit to obtain the same cluster density as observed in the target redshift range and constructed  superstructures with a linking length of $38.5~ h_{70}^{-1}$ Mpc.
In a volume that is $\sim 5.4$ times larger than our survey we find 23 superstructures.
Since in our observational study the Hercules supercluster has less than 20 clusters inside the target volume, we should expect about $21.6 \pm 4.6$ superstructures in the simulation in good agreement with the observations.
Fig.~\ref{fig:comp_multi} compares the multiplicity functions (histograms of the number of members) for the observed and simulated superstructures, showing similar distributions. 
The two functions are normalised by the ratio of the two survey volumes.
Thus the superstructures we found are expected in conventional $\Lambda$CDM models in the generally adopted concordance cosmology \citep{Pla2016a}.

\begin{figure} 
	\centering
	\includegraphics[width=0.45\textwidth]{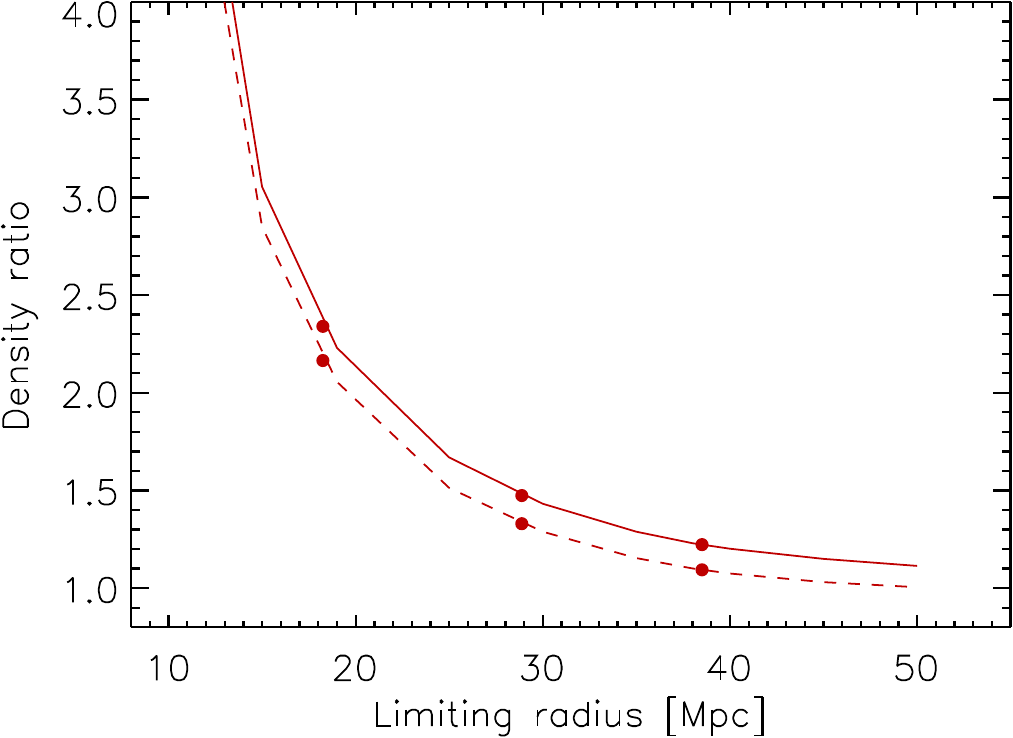}
	\caption{
    Ratio of the matter density around superstructures to the mean matter density as a function of the limiting radius (solid curve) compared to that of field clusters (dashed line). 
    The radii for which the values are listed in Table\ref{tab:densities} are marked with solid points.
 }
\label{fig:densdist} 
\end{figure} 

\begin{figure} % Do NOT use \begin{figure*}
	\centering
	\includegraphics[width=0.45\textwidth]{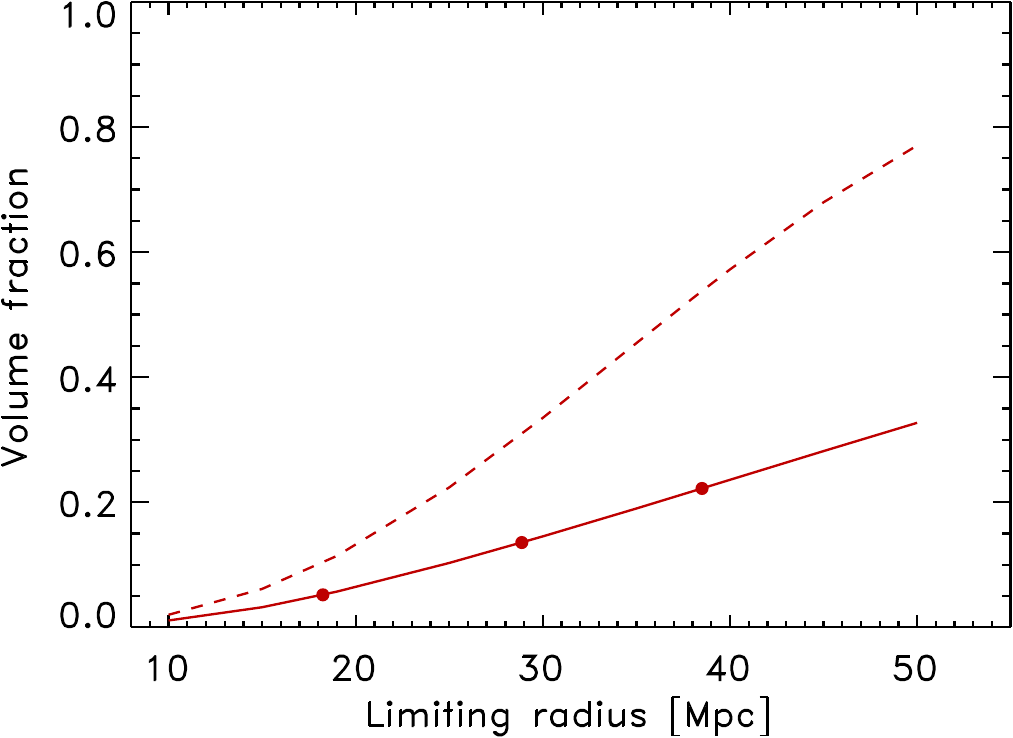}
	\caption{
    Volume fraction filled by superstructures (solid line) and field clusters with their halos (dashed line) as a function of limiting radius. 
 }
\label{fig:volfill} 
\end{figure} 

The simulation also allows us to study the connection of the cluster overdensities with the underlying large-scale matter distribution. 
Therefore we determined the relative matter density in the superstructures as a function of the distance to the nearest cluster member, the limiting radius, in an analogous fashion to the study of the galaxy density distribution in the previous section. 
We also performed the same calculations for the clusters outside the superstructures and show the results in Fig.~\ref{fig:densdist}.
We note again a difference in the density distribution of the superstructures
compared to the sample of field clusters, with the densities in the halos of
superstructures being higher. 
But this difference is less pronounced than that for the galaxy density distribution.

\begin{table} % Do NOT use \begin{table*}
	%\centering
	% Captions go above tables
	\caption{Overdensties and volume filling factors of the
superstructures in simulations and observations}
%\medskip
	\label{tab:densities} % give each table a logical label name
\begin{tabular}{@{}lrrr@{}}
\hline
{\rm Region} &  {\rm  18.75~Mpc}& {\rm 28.875~Mpc}  & {\rm 38.5~Mpc} \\ 
\hline
{\rm Superstructures} & & &\\
\hline
{\rm Matter~overdensity$^a$}    & 1.3    & 0.47    & 0.23 \\
{\rm Galaxy~overdensity$^b$}  & 3.6    & 1.8     & 1.2 \\
{\rm Cluster~overdensity$^c$} & 8.5    & 3.3     & 1.6 \\
{\rm Cluster~overdensity$^a$} & 5.9    & 1.6     & 0.6  \\
{\rm Volume~fraction$^c$}     & 4.7\%  & 10.8\%  & 17.6\% \\
{\rm Volume~fraction$^a$}     & 5.2\%  & 13.5\% & 22\% \\ 
\hline
{\rm Clusters~outside} & & & \\
\hline
{\rm Matter~overdensity$^a$}    & 1.2  & 0.33 & 0.11 \\
{\rm Galaxy~overdensity$^b$}  & 1.7  & 0.7  & 0.3  \\
\hline
\end{tabular}	
\smallskip

{\bf Notes:} Overdensities are defined as $\Delta = (\rho - <\rho> )/ \rho$.\\
$^{a)}$ determined in the Millenium simulations.
$^{b)}$ determined with the 2MASS sample.
$^{c)}$ for CLASSIX galaxy clusters. 
\end{table}

In Table~\ref{tab:densities} we list the matter, cluster, and galaxy overdensities as well as volume fractions of the superstructures for three selected limiting radii, 18.25, 28.875, and 38.5 Mpc, which correspond to half, 3/4, and full linking length.
The values for the radius 28.875 Mpc can be directly compared to the values
estimated for Table~\ref{tab:Tab1}. 
The matter overdensity in the simulated superstructures using this fiducial radius is $\Delta_{DM} = 0.5$, while that of the environment of all non-member clusters is $\Delta_{DM} = 0.3$.
For the observed superstructures we estimated matter overdensities in the range $0.3$ to $1.3$.
For a better comparison we determined the overdensity values for all observed superstructures (with some overlapping volume) and find $\Delta_{DM} = 0.83$ for superstructures and $\Delta_{DM} = 0.43$ for all other clusters. 
These numbers are somewhat higher than those in the simulations, but still comparable.
If we compare these numbers with the overdensities of 2MASS galaxies, for which we do not expect a large bias, we see that the galaxy overdensities in superstructures are with a value of $\Delta_{gal} = 1.8$  higher than the estimated matter overdensity found in simulations.  
The value of $\Delta_{gal} = 0.7$ for the environment of the non-member clusters is, however, closer to the matter density in these regions.
Also for the cluster overdensities in superstructures we find somewhat larger values for the observations than the simulations, possibly indicating a more compact configuration of observed compared to simulated superstructures. 
This may also partly explain the higher values of the matter overdensities.

The significantly higher matter and galaxy overdensities in the surroundings of clusters in superstructures compared to clusters in the field is a signature that superstructure environments are special places in the Universe. 
While this is a first impression from the current data, it should be an important goal of future, much larger surveys, to investigate these special properties of superstructures at high precision. In Fig.~\ref{fig:volfill} we show the volume fraction occupied by member and non-member clusters as a function of limiting radius.

In summary we conclude from the observations that superstructures are a major component of the Universe, containing 45\% of the clusters, $\sim 30\%$ of the galaxies, $\sim 25\%$ of the matter and occupy $\sim 13\%$ of the volume. 
In comparison, more typical superclusters (including cluster pairs) with overdensities of $\sim 7$ contain about half of the clusters in a space fraction of $\sim 10\%$ \citep{Cho2014}.

% ======================================================================
\section{Discussion of the results}
% ======================================================================

\begin{figure} % Do NOT use \begin{figure*}
	\centering
	\includegraphics[width=0.45\textwidth]{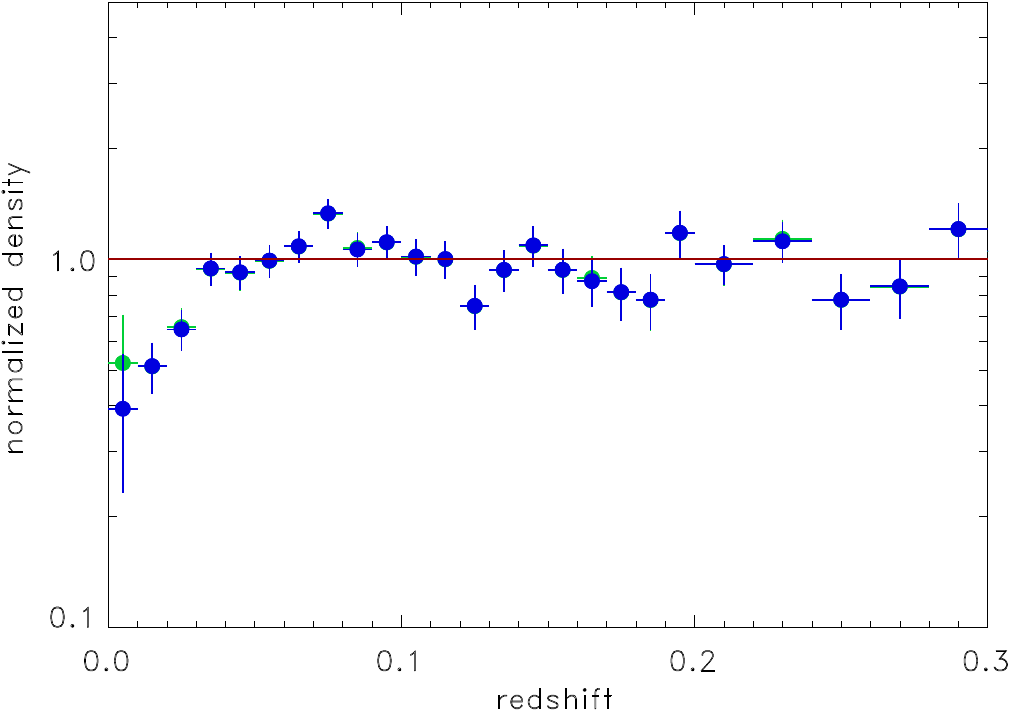}
	\caption{
    Cluster density distribution as a function of redshift for the {\sf CLASSIX} galaxy clusters covering the sky at $|b_{II}| \ge 20^o$ for a minimum luminosity of $10^{42}$ erg s$^{-1}$ (0.1 - 2.4 keV). 
    The density distribution has been normalised by the local mean cluster density based on the average cluster luminosity function \citep{Boe2020}.
    The green open diamond shows the result if the region of the Virgo cluster is not excluded from the analysis.
 }
\label{fig:raddist} % give each figure a logical label name
\end{figure}

With our search criteria we found five superstructures in the redshift region $z = 0.03$ to $0.06$.
With the same criteria we find only one such structure in the local volume at $z \le 0.03$. 
The target volume of this study is eight times larger than the local volume. 
While the number of five superstructures falls slightly short in comparison, the numbers of involved clusters are with 155 (185) for the target volume and 19 (22) for the local volume well comparable. 
Here the numbers relate to the clusters in the target volume, while those in brackets refer to all member clusters of the superstructures. 
Thus the essential difference is the higher degree of clustering in the target volume. 
It could be a result of the higher cluster density in the redshift range $z = 0.03 - 0.06$. 
This is illustrated in Fig.~\ref{fig:raddist}, where we show the redshift distribution of the density of the {\sf CLASSIX} galaxy clusters normalised to the global average \citep{Boe2020}. 
The local volume is characterised by an underdensity while the target volume has a higher density which is more comparable to the large-scale average density.

Another large cosmological volume analysed for the presence of large structures is the Sloan Digital Sky Survey. 
The largest structure found there is the Sloan Great Wall, described by e.g. \citet{Ein2016} with a length of about 328 Mpc, which in their analysis breaks up into 3 smaller superclusters. 
Studies of the topology of the large-scale structure in cosmological simulations \citep{Cau2014,Lib2018} have shown that large filaments with a size $\ge 100~ h_{100}^{-1}$~Mpc dominate the cosmic web since a redshift of about 2, carrying up to 50\% of the total mass in the Universe at present. 
We can identify these structures with the superstructures found in our survey.

In the previous section we explored the cluster, galaxy, and the matter density in superstructures and outside, which are summarised in Table~\ref{tab:densities}.
Among these results the difference in the galaxy density around field clusters and members of superstructures is remarkable. 
To make it clear, this is not due to the fact, that at a certain distance from the cluster the density can be high, since it overlaps with another cluster whose centre is closer, because we have taken as the relevant radius always the closest distance to any cluster. 
Also this is done in three dimensions, thus there is no overlap in projection.
Therefore this result can be taken as a genuine property of the superstructures and its reason needs to be further explored.

\begin{figure} % Do NOT use \begin{figure*}
	\centering
	\includegraphics[width=0.45\textwidth]{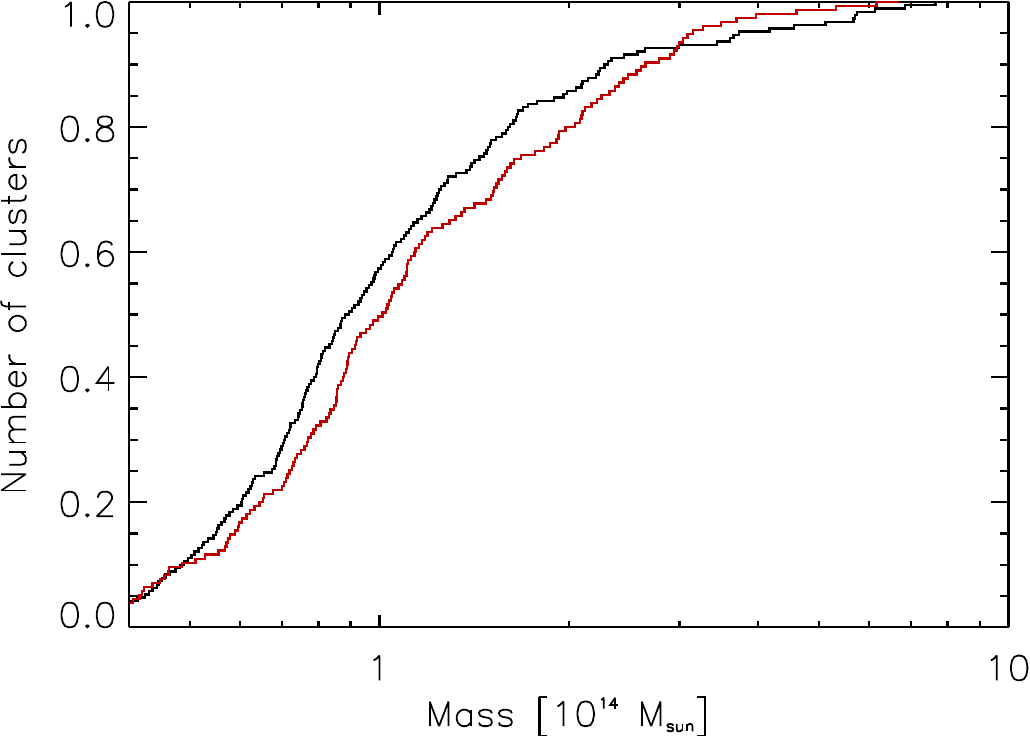}
	\caption{Normalised cumulative mass function of clusters in superstructures (red) and field clusters (black).
 }
\label{fig:clpop} 
\end{figure} 

One cause of a higher galaxy density in superstructures could in principle be that the cluster population of superstructures involves more massive clusters than that of the field. 
We explore this in Fig.~\ref{fig:clpop}, where we compare the cumulative cluster mass functions of both populations. 
The two curves are quite similar and a Kolmogorov-Smirnov test indicates with a probability value of 0.127 that the difference between the two curves is not significant.
The effect is therefore too small to be the cause of the higher galaxy density around superstructures compared to the field.
In the study of classical superclusters with much higher overdensities and the constraint that they are gravitationally bound, we found a significant result for a top heavy mass function in superclusters compared to the field \citep{Cho2015}.
Thus, at least at higher overdensities the difference becomes evident.

% ======================================================================
\section{Imprint of the superstructures on cosmological observations}
% ======================================================================

These large structures leave their imprint on cosmological observations. 
In the following we discuss a few of these aspects relevant to our findings.

% ======================================================================
\subsection{Peculiar velocity of the Local Group and effects on the Hubble flow.}
% ======================================================================

The local large-scale structure is the cause of large-scale streaming motions, detected, for example, in the peculiar velocities of galaxies. 
Of special importance in this context is an understanding of the origin of the motion of the Local Group, $\vec{v}_{LG}$, with respect to the cosmic microwave background (CMB) reference frame \citep{Yah1980,Dav1982,Str1992,Row2000}. 
This peculiar motion can be detected as a Doppler effect in the CMB radiation. 
From this radiation dipole a velocity of $v_{LG} = 276 \pm 22$ km s$^{-1}$ and a direction of $(l,b) = (276^{\circ} \pm 3^{\circ}, 30^{\circ} \pm 3^{\circ})$ was determined \citep{Kog1993}. 
This motion should originate from the gravitational pull of the surrounding large-scale structure.
To provide a consistent picture of the observed CMB dipole and the prediction of $v_{LG}$ from assessment of the local large-scale structure is therefore an important task of observational cosmology. 
The analysis of the large-scale structure seen in the galaxy distribution of comprehensive redshift surveys \citep{Yah1980,Dav1982,Str1992,Row2000} has not yet provided conclusive and consistent results. 
As shown in a systematic simulation study by \citet{Nus2014} the volume needed to cover the relevant cosmic structures requires a convergence depth of about 250 Mpc, which is not reached by current galaxy redshift surveys.
In another approach, galaxy clusters, which cover a sufficiently large volume, have been used to reconstruct the inhomogeneous mass distribution and its gravitational influence on the Local Group, including cluster samples from optical as well as X-ray surveys \citep{Sca1991,Pli1991,Pli1998}. 
These studies agreed on finding a first preliminary plateau in the prediction of $v_{LG}$ at a survey depth of about 70 Mpc.
This is followed, however, by a second increase in the predicted velocity with increasing survey depth which saturates at about 200 - 250 Mpc. 
The final value is consistent with the velocity inferred from the CMB dipole within the measurement errors. 
The height of the first plateau is about one half to two thirds of the final value. 
The main structure responsible for this influence from the distance can be located in the Shapley supercluster.

From these results it is clear that the superstructures reported in this paper are important for our understanding of the peculiar motion of the Local Group. 
This result is further supported by a study of the distribution of galaxy peculiar motions in the local Universe, which provides a complementary analysis of the large-scale structure to the direct study of the density distribution. 
Results from Tully and coworkers \citep{Tul2016} using mostly Tully-Fisher distances and redshifts for the peculiar velocity measurements, show that the most important attractor on their largest scales is the Shapley supercluster. 
The bulk flow, in which the Local Group is embedded, is caused to a large degree by this mass concentration as well as a major underdensity on the opposite side, termed ''Cold Spot Repeller'' \citep{Hof2015,Cou2017}.

A simple estimate of the effect on the measurement of the Hubble constant in the presence of such mass concentrations can be obtained as follows. 
The expected streaming motions towards the attractors amount to several 100 km s$^{-1}$.
Taking for example a region in front of the Shapley supercluster in the distance range $cz = 10\,000$ to $15\,000$ km s$^{-1}$ leads to expected deviations of the local values of $H_0$ of one to a few percent.
While this will be difficult to detect in the present data, it should be observable with the increasing precision and statistics of $H_0$ measurements.

% ======================================================================
\subsection{ISW effect}
% ======================================================================

The CMB, which originates mostly from the Universe at a redshift around 1100, providing valuable information about the early cosmic epochs \citep{Pla2016a}, shows another signature from the nearby large-scale structure through the Integrated Sachs-Wolfe (ISW) effect \citep{Sac1967}. 
This is caused when CMB photons pass through the gravitational potential of the evolving large-scale structure. 
In the context of $\Lambda$CDM cosmology the potentials of superstructures of the kind described here become shallower in the course of time since the density decrease due to cosmic expansion can be faster than the overdensity growth. The photons gain therefore more potential energy falling into the potential than they loose when coming out, which leads to a temperature increase of the radiation. 
Theoretical modelling suggests that such an effect from the superstructures can amount to several $\mu$K and can in principle be detected, as shown in the next section.

% ======================================================================
\subsubsection{Modeling the ISW effect}
% ======================================================================

To get an idea of the magnitude of the ISW effect we expect for the superstructures, we studied four simple, spherically symmetric superstructure models. 
For model 1 and 2 we assume a top-hat matter distribution, with a constant overdensity inside a radius of 40 Mpc and no overdensity outside. 
We trace the potential of this structure out to 80 Mpc.
For model 3 and 4 we assume a matter density profile described by a power law, $\rho \propto r^{-1.8}$, also traced up to 80 Mpc. 
The superstructure mass inside 40 Mpc is assumed to be $2 - 3\times 10^{16}$ M$_{\odot}$. 
More details on the model parameters are given in Table~\ref{tab:ISW}.      

The ISW effect is then obtained by (e.g. \citealt{Cri1996,Cai2010}):

\begin{equation}
\Delta T = {2 \over c^2}~ T_{cmb} \int_{t_i}^{t_0}{ \dot{\phi}~ dt} ,
\end{equation}

where $\phi$ is the time variable gravitational potential formulated in a comoving frame.
This potential is calculated from the Poisson equation taking into account the evolution of the overdensity with time and the expansion of the Universe. 
Since these structures are evolving out of the linear regime, we use a numerically calculated non-linear growth function. 
In the model calculation we take a line of sight through the centre of the superstructure, which is justified below.
The results of the calculations are given in Table~\ref{tab:ISW}.
We find temperature enhancements of about  3 to 5 $\mu K$.

\begin{table} % Do NOT use \begin{table*}
	\centering
	\caption{
    Superstructure models to estimate the magnitude of the ISW effect.}
\medskip
\label{tab:ISW} % give each table a logical label name
\begin{tabular}{@{}lrrlr@{}}
\hline
 model &  {\rm mass} & Delta & {\rm shape} & {\rm signal}\\ 
\hline
1 & $ 2\times 10^{16}$   &  2   & {\rm top~hat}             & 2.9$\mu$K \\
2 & $ 2.5\times 10^{16}$ &  2.5 & {\rm top~hat}             & 4.2$\mu$K \\
3 & $ 2\times 10^{16}$   &  2   & $\rho \propto r^{-1.8}$ {\rm profile} & 3.0$\mu$K \\
4 & $ 3\times 10^{16}$   &  3   & $\rho \propto r^{-1.8}$ {\rm profile} & 5.3$\mu$K \\
\hline
\end{tabular}	
\smallskip

{\bf Notes:} The mass quoted in column 2 is the mass inside a radius of 40 Mpc in units of M$_{\odot}$. 
    The top hat model extends to a radius of 40 Mpc and the $\rho^{-1.8}$ profile to a radius of 80 Mpc.
\end{table}

\begin{figure} 
	\centering
	\includegraphics[width=0.45\textwidth]{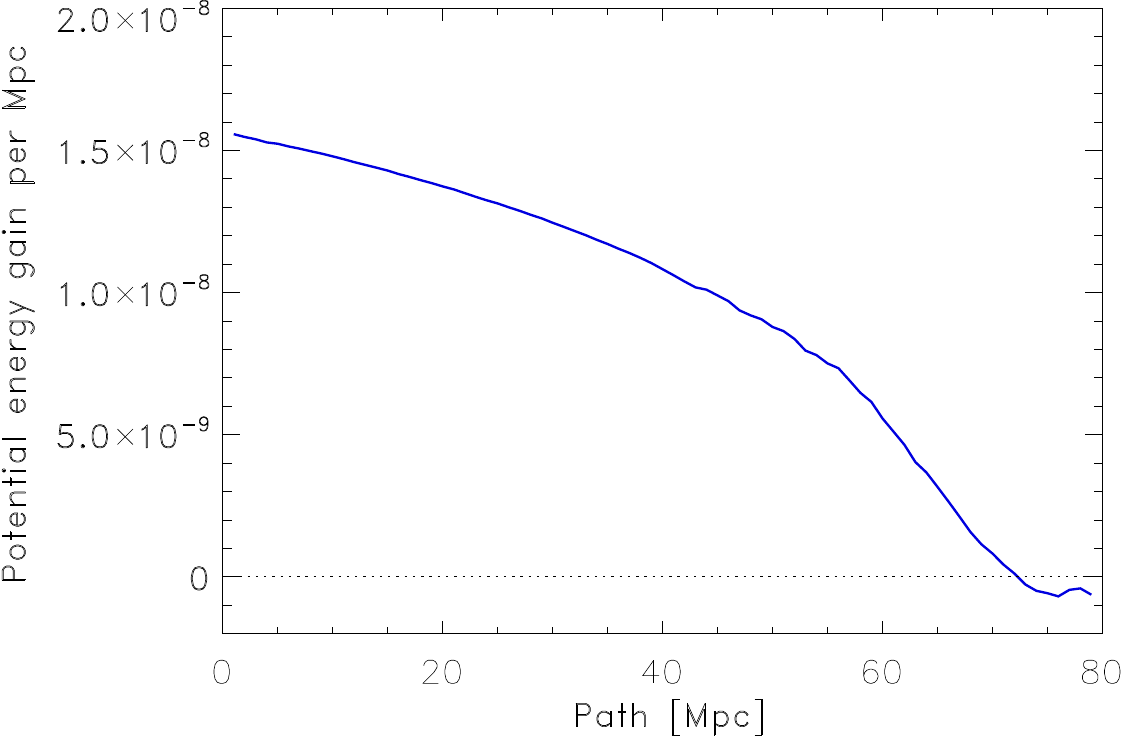}
	\caption{
    Differential potential energy gain comparing corresponding 1 Mpc intervals in the incoming and outgoing path of the photons for model 3. 
    The supercluster centre corresponds to the path location at 80 Mpc.
 }
\label{fig:ISW_signal} % give each figure a logical label name
\end{figure} 

We have also investigated which part of the light path contributes most to the energy gain of the CMB photons. 
For this we compare the potential difference at two points in the incoming and outgoing light path at the same distance from the superstructure centre.
We calculate the difference of the positive and negative potential change for each step of 1 Mpc.
The result of this comparison is shown for the case of model 3 in Fig.~\ref{fig:ISW_signal}. 
In the very centre, at 80 Mpc in the plot, we see an energy loss, due to the fast non-linear structure growth. 
What is important here, is that it is not the centre which contributes most to the energy gain, but the outskirts provide a substantial contribution. 
Therefore also light rays which do not go through the centre will have substantial ISW energy shifts. 
The superstructures will make an imprint on the CMB over their entire extent. 
The magnitude of the effect is comparable to the cosmic density fluctuations in the CMB.

% ======================================================================
\subsubsection{Searching for the ISW effect in the Planck data}
% ======================================================================

To search for the ISW effect we used the Planck CMB maps from the third data release \citep{Pla2016b}, which have been cleaned from all foregrounds with the commander, nilc, and smica methods. 
The data sets of all three maps include masks which mark the regions containing serious contamination, mainly near the 
Galactic plane. 
These maps were used to exclude these regions in our correlation analysis. 
We used the Healpix maps at highest resolution with $N = 1024$. 

\begin{figure} 
	\centering
	\includegraphics[width=0.45\textwidth]{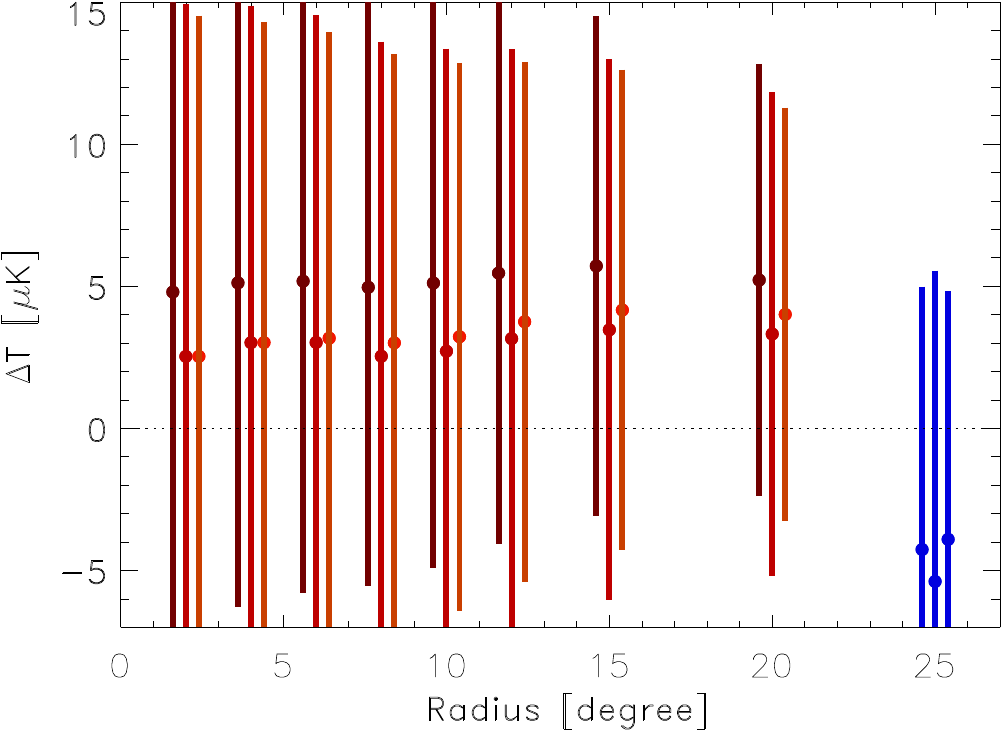}
	\caption{
        Integrated Sachs-Wolfe temperature increment in concentric circles around each member cluster of the superstructures. 
    The three data points show the signal from the commander (left offset), nilc (middle) and smica (right offset) analysis codes. 
    The blue points show the signal from the sky region further away from any member cluster than 20 degrees.
    The error bars indicate 1$\sigma$ uncertainties.
 }
\label{fig:Planck} % give each figure a logical label name
\end{figure}

We first determined the correlation of the superstructures and the CMB in a differential way as function of radius. We sorted all sky pixels into radial annuli according to the closest distance to any member of the superstructures. 
The signal was then averaged over all pixels in the bin.
We found a signal out to a distance of 15 degrees, which decreased to about half at 20 degrees and then vanished. 
In a second step we probed the cumulative signal by sorting the sky pixels into circular apertures according to the shortest distances to any member cluster, ranging from 2 to 20 degrees.
The results are shown in Fig.~\ref{fig:Planck}. 
We note that the highest significance is found at a radius of 15 to 20 degrees. 
The blue data points in the figure give the mean CMB temperature in the rest of the sky outside a distance of 20 degrees to any cluster member.
The noise in the signal comes mainly from the statistical fluctuations of the primordial CMB. 
To determine this noise for our case, we placed the same aperture masks of the five superstructures on 100 simulations of the CMB for each of the three methods. 
The simulated data were obtained from the Wiki page of the Planck science archive
\footnote{https://wiki.cosmos.esa.int/planck-legacy-archive/index.php/CMB\_maps}.
The magnitude of the signal is exactly in the range predicted by the above described models. 
We note, however, that the variance of the original CMB fluctuations is of similar size as the ISW signal. Therefore the detection has unfortunately a low significance of less than $1\sigma$. 

Signatures of the ISW effect have been detected earlier from statistical studies of tracer - CMB correlation functions, e.g. \citep{Gia2012,Nad2016}, and the stacking of many superclusters and voids \citep{Gra2008} with signals hardly exceeding 3$\sigma$.
It is generally difficult to separate the ISW signal from the CMB fluctuations even for larger survey volumes.

There is some leverage to improve the significance of the results in the future, on one hand with a tailored matched filter detection method, which requires a better understanding of the mass distribution in the superstructures than we have at present. 
On the other hand an improvement can be achieved, by including CMB polarisation data on large-scale, which are not available at present, as suggested by \citep{Fro2009,Pla2016c}.
Including the superstructures at higher redshift could also help to improve the signal.
A more precise knowledge of the ISW effect would in turn also allow to clean the CMB maps from this foreground modification. 

% ======================================================================
\section{Conclusion and outlook}
% ======================================================================

In this study we characterised six large superstructures with more than 20 {\sf CLASSIX} galaxy cluster members in the nearby Universe out to a redshift of $z = 0.06$. 
The only superstructure found in the local Universe at $z \le 0.03$ with these properties, the Perseus-Pisces supercluster, has been described earlier \citep{Boe2021b}. 
For the present superstructure construction we only included clusters with an estimated minimal mass of $\ge 10^{13}$ M$_{\odot}$ (this is only important for $z \le 0.03$, since for higher redshifts lower mass clusters are not included due to
the flux limit of the sample). 
The largest structure ever detected, Quipu, with a length of 428 Mpc, is located in the redshift region $z= 0.027 - 0.065$.
This structure is clearly apparent in a map of the cluster or the galaxy distribution in the redshift range $z =0.03 - 0.06$.
The highest concentrated mass overdensity in the target volume of our study is found in the Shapley supercluster. 
This structure has been made responsible for a large part of the gravitational pull that causes the peculiar motion of the Local Group with respect to the CMB frame.

We showed that the galaxy density in the environment of galaxy clusters in superstructures is significantly higher than around field clusters. 
This characterises superstructures as special astrophysical environments. 
We also showed that these structures should leave a signature in the CMB through the integrated Sachs-Wolf effect, comparable to the cosmic density fluctuations in the background. 
Searching for this effect we could only recover a signal with a significance close to 1$\sigma$, but with the expected strength.

Interesting follow-up research on our findings includes, for example, studies of the influence of these environments 
on the galaxy population and evolution. 
In an earlier paper Giovanelli and coworkers \citep{Gio1986} show the galaxy distribution of different galaxy types in the Perseus-Pisces supercluster. 
In their Fig.~6b we can see a very sharp image of the filamentary structure of the supercluster on 100 Mpc scales traced by early type galaxies. 
This is very different from the outline of the supercluster by other galaxy types. 
Detailed studies of the larger structures described here may show a similar effects over even larger scales.
Similarly the dense environment and the large fraction of the volume heated by accretion shocks in superstructures can, for example, be expected to have an imprint on the entropy structure of galaxy groups. 
Thus X-ray studies of galaxy groups in superstructure environments compared to observations in the field could shed new light on the formation of the intra-cluster medium.   

In the future cosmic evolution, these superstructures are bound to break up into several collapsing units. 
They are thus transient configurations. 
But at present they are special physical entities with characteristic properties and special cosmic environments deserving special attention.

\section{Data availability}

The full cluster catalogue shown in the appendix is only available in electronic form at CDS via anonymous ftp to cdsarc.u-strasbg.fr (130.79.128.5) or via http://cdsweb.u-strasbg.fr/cgi-bin/qcat?J/A+A/..

% ======================================================================
\begin{acknowledgements}
% ======================================================================
HB acknowledges support from the Deutsche Forschungsgemeinschaft through the Munich Excellence Cluster ''Origins''. 
RKK thanks the South African National Science Foundation for support.
\end{acknowledgements}

\bibliographystyle{aa} % style aa.bst
\bibliography{Superstruc} % .bib file

\begin{thebibliography}{67}
\expandafter\ifx\csname natexlab\endcsname\relax\def\natexlab#1{#1}\fi

\bibitem[{{Balaguera-Antol{\'\i}nez} {et~al.}(2011){Balaguera-Antol{\'\i}nez}, {S{\'a}nchez}, {B{\"o}hringer}, {Collins}, {Guzzo}, \& {Phleps}}]{Bal2011}
{Balaguera-Antol{\'\i}nez}, A., {S{\'a}nchez}, A.~G., {B{\"o}hringer}, H., {et~al.} 2011, \mnras, 413, 386

\bibitem[{{Barmby} \& {Huchra}(1998)}]{Bar1998}
{Barmby}, P. \& {Huchra}, J.~P. 1998, \aj, 115, 6

\bibitem[{{Bilicki} {et~al.}(2014){Bilicki}, {Jarrett}, {Peacock}, {Cluver}, \& {Steward}}]{Bil2014}
{Bilicki}, M., {Jarrett}, T.~H., {Peacock}, J.~A., {Cluver}, M.~E., \& {Steward}, L. 2014, \apjs, 210, 9

\bibitem[{{B{\"o}hringer} \& {Chon}(2021)}]{Boe2021c}
{B{\"o}hringer}, H. \& {Chon}, G. 2021, \aap, 656, A144

\bibitem[{{B{\"o}hringer} {et~al.}(2020){B{\"o}hringer}, {Chon}, \& {Collins}}]{Boe2020}
{B{\"o}hringer}, H., {Chon}, G., \& {Collins}, C.~A. 2020, \aap, 633, A19

\bibitem[{{B{\"o}hringer} {et~al.}(2013){B{\"o}hringer}, {Chon}, {Collins}, {Guzzo}, {Nowak}, \& {Bobrovskyi}}]{Boe2013}
{B{\"o}hringer}, H., {Chon}, G., {Collins}, C.~A., {et~al.} 2013, \aap, 555, A30

\bibitem[{{B{\"o}hringer} {et~al.}(2017){B{\"o}hringer}, {Chon}, {Retzlaff}, {Tr{\"u}mper}, {Meisenheimer}, \& {Schartel}}]{Boe2017}
{B{\"o}hringer}, H., {Chon}, G., {Retzlaff}, J., {et~al.} 2017, \aj, 153, 220

\bibitem[{{B{\"o}hringer} {et~al.}(2021{\natexlab{a}}){B{\"o}hringer}, {Chon}, \& {Tr{\"u}mper}}]{Boe2021a}
{B{\"o}hringer}, H., {Chon}, G., \& {Tr{\"u}mper}, J. 2021{\natexlab{a}}, \aap, 651, A15

\bibitem[{{B{\"o}hringer} {et~al.}(2021{\natexlab{b}}){B{\"o}hringer}, {Chon}, \& {Tr{\"u}mper}}]{Boe2021b}
{B{\"o}hringer}, H., {Chon}, G., \& {Tr{\"u}mper}, J. 2021{\natexlab{b}}, \aap, 651, A16

\bibitem[{{B{\"o}hringer} {et~al.}(2012){B{\"o}hringer}, {Dolag}, \& {Chon}}]{Boe2012}
{B{\"o}hringer}, H., {Dolag}, K., \& {Chon}, G. 2012, \aap, 539, A120

\bibitem[{{B{\"o}hringer} {et~al.}(2004){B{\"o}hringer}, {Schuecker}, {Guzzo}, {Collins}, {Voges}, {Cruddace}, {Ortiz-Gil}, {Chincarini}, {De Grandi}, {Edge}, {MacGillivray}, {Neumann}, {Schindler}, \& {Shaver}}]{Boe2004}
{B{\"o}hringer}, H., {Schuecker}, P., {Guzzo}, L., {et~al.} 2004, \aap, 425, 367

\bibitem[{{B{\"o}hringer} {et~al.}(2000){B{\"o}hringer}, {Voges}, {Huchra}, {McLean}, {Giacconi}, {Rosati}, {Burg}, {Mader}, {Schuecker}, {Simi{\c{c}}}, {Komossa}, {Reiprich}, {Retzlaff}, \& {Tr{\"u}mper}}]{Boe2000}
{B{\"o}hringer}, H., {Voges}, W., {Huchra}, J.~P., {et~al.} 2000, \apjs, 129, 435

\bibitem[{{Bulbul} {et~al.}(2024){Bulbul}, {Liu}, {Kluge}, {Zhang}, {Sanders}, {Bahar}, {Ghirardini}, {Artis}, {Seppi}, {Garrel}, {Ramos-Ceja}, {Comparat}, {Balzer}, {B{\"o}ckmann}, {Br{\"u}ggen}, {Clerc}, {Dennerl}, {Dolag}, {Freyberg}, {Grandis}, {Gruen}, {Kleinebreil}, {Krippendorf}, {Lamer}, {Merloni}, {Migkas}, {Nandra}, {Pacaud}, {Predehl}, {Reiprich}, {Schrabback}, {Veronica}, {Weller}, \& {Zelmer}}]{Bul2024}
{Bulbul}, E., {Liu}, A., {Kluge}, M., {et~al.} 2024, \aap, 685, A106

\bibitem[{{Cai} {et~al.}(2010){Cai}, {Cole}, {Jenkins}, \& {Frenk}}]{Cai2010}
{Cai}, Y.-C., {Cole}, S., {Jenkins}, A., \& {Frenk}, C.~S. 2010, \mnras, 407, 201

\bibitem[{{Cautun} {et~al.}(2014){Cautun}, {van de Weygaert}, {Jones}, \& {Frenk}}]{Cau2014}
{Cautun}, M., {van de Weygaert}, R., {Jones}, B. J.~T., \& {Frenk}, C.~S. 2014, \mnras, 441, 2923

\bibitem[{{Chon} \& {B{\"o}hringer}(2012)}]{Cho2012}
{Chon}, G. \& {B{\"o}hringer}, H. 2012, \aap, 538, A35

\bibitem[{{Chon} {et~al.}(2014){Chon}, {B{\"o}hringer}, {Collins}, \& {Krause}}]{Cho2014}
{Chon}, G., {B{\"o}hringer}, H., {Collins}, C.~A., \& {Krause}, M. 2014, \aap, 567, A144

\bibitem[{{Chon} {et~al.}(2015){Chon}, {B{\"o}hringer}, \& {Zaroubi}}]{Cho2015}
{Chon}, G., {B{\"o}hringer}, H., \& {Zaroubi}, S. 2015, \aap, 575, L14

\bibitem[{{Courtois} {et~al.}(2023){Courtois}, {Dupuy}, {Guinet}, {Baulieu}, {Ruppin}, \& {Brenas}}]{Cou2023}
{Courtois}, H.~M., {Dupuy}, A., {Guinet}, D., {et~al.} 2023, \aap, 670, L15

\bibitem[{{Courtois} {et~al.}(2019){Courtois}, {Kraan-Korteweg}, {Dupuy}, {Graziani}, \& {Libeskind}}]{Cou2019}
{Courtois}, H.~M., {Kraan-Korteweg}, R.~C., {Dupuy}, A., {Graziani}, R., \& {Libeskind}, N.~I. 2019, \mnras, 490, L57

\bibitem[{{Courtois} {et~al.}(2017){Courtois}, {Tully}, {Hoffman}, {Pomar{\`e}de}, {Graziani}, \& {Dupuy}}]{Cou2017}
{Courtois}, H.~M., {Tully}, R.~B., {Hoffman}, Y., {et~al.} 2017, \apjl, 847, L6

\bibitem[{{Crittenden} {et~al.}(1996){Crittenden}, {Boughn}, \& {Turok}}]{Cri1996}
{Crittenden}, R., {Boughn}, S., \& {Turok}, N. 1996, in American Astronomical Society Meeting Abstracts, Vol. 189, American Astronomical Society Meeting Abstracts, 51.04

\bibitem[{{Davis} \& {Huchra}(1982)}]{Dav1982}
{Davis}, M. \& {Huchra}, J. 1982, \apj, 254, 437

\bibitem[{{de Lapparent} {et~al.}(1986){de Lapparent}, {Geller}, \& {Huchra}}]{Lap1986}
{de Lapparent}, V., {Geller}, M.~J., \& {Huchra}, J.~P. 1986, \apjl, 302, L1

\bibitem[{{Dolag} {et~al.}(2023){Dolag}, {Sorce}, {Pilipenko}, {Hern{\'a}ndez-Mart{\'\i}nez}, {Valentini}, {Gottl{\"o}ber}, {Aghanim}, \& {Khabibullin}}]{Dol2023}
{Dolag}, K., {Sorce}, J.~G., {Pilipenko}, S., {et~al.} 2023, \aap, 677, A169

\bibitem[{{Einasto} {et~al.}(2016){Einasto}, {Lietzen}, {Gramann}, {Tempel}, {Saar}, {Liivam{\"a}gi}, {Hein{\"a}m{\"a}ki}, {Nurmi}, \& {Einasto}}]{Ein2016}
{Einasto}, M., {Lietzen}, H., {Gramann}, M., {et~al.} 2016, \aap, 595, A70

\bibitem[{{Frommert} \& {En{\ss}lin}(2009)}]{Fro2009}
{Frommert}, M. \& {En{\ss}lin}, T.~A. 2009, \mnras, 395, 1837

\bibitem[{{Fry} \& {Peebles}(1980)}]{Fry1980}
{Fry}, J.~N. \& {Peebles}, P.~J.~E. 1980, \apj, 238, 785

\bibitem[{{Giannantonio} {et~al.}(2012){Giannantonio}, {Crittenden}, {Nichol}, \& {Ross}}]{Gia2012}
{Giannantonio}, T., {Crittenden}, R., {Nichol}, R., \& {Ross}, A.~J. 2012, \mnras, 426, 2581

\bibitem[{{Giovanelli} {et~al.}(1986){Giovanelli}, {Haynes}, \& {Chincarini}}]{Gio1986}
{Giovanelli}, R., {Haynes}, M.~P., \& {Chincarini}, G.~L. 1986, \apj, 300, 77

\bibitem[{{Granett} {et~al.}(2008){Granett}, {Neyrinck}, \& {Szapudi}}]{Gra2008}
{Granett}, B.~R., {Neyrinck}, M.~C., \& {Szapudi}, I. 2008, \apjl, 683, L99

\bibitem[{{Groth} \& {Peebles}(1977)}]{Gro1977}
{Groth}, E.~J. \& {Peebles}, P.~J.~E. 1977, \apj, 217, 385

\bibitem[{{Guzzo} {et~al.}(2009){Guzzo}, {Schuecker}, {B{\"o}hringer}, {Collins}, {Ortiz-Gil}, {de Grandi}, {Edge}, {Neumann}, {Schindler}, {Altucci}, \& {Shaver}}]{Guz2009}
{Guzzo}, L., {Schuecker}, P., {B{\"o}hringer}, H., {et~al.} 2009, \aap, 499, 357

\bibitem[{{Hoffman} {et~al.}(2015){Hoffman}, {Courtois}, \& {Tully}}]{Hof2015}
{Hoffman}, Y., {Courtois}, H.~M., \& {Tully}, R.~B. 2015, \mnras, 449, 4494

\bibitem[{{Huchra} {et~al.}(2012){Huchra}, {Macri}, {Masters}, {Jarrett}, {Berlind}, {Calkins}, {Crook}, {Cutri}, {Erdo{\v{g}}du}, {Falco}, {George}, {Hutcheson}, {Lahav}, {Mader}, {Mink}, {Martimbeau}, {Schneider}, {Skrutskie}, {Tokarz}, \& {Westover}}]{Huc2012}
{Huchra}, J.~P., {Macri}, L.~M., {Masters}, K.~L., {et~al.} 2012, \apjs, 199, 26

\bibitem[{{J{\~o}eveer} {et~al.}(1978){J{\~o}eveer}, {Einasto}, \& {Tago}}]{Joe1978}
{J{\~o}eveer}, M., {Einasto}, J., \& {Tago}, E. 1978, \mnras, 185, 357

\bibitem[{{Jasche} \& {Lavaux}(2019)}]{Jas2019}
{Jasche}, J. \& {Lavaux}, G. 2019, \aap, 625, A64

\bibitem[{{Kogut} {et~al.}(1993){Kogut}, {Lineweaver}, {Smoot}, {Bennett}, {Banday}, {Boggess}, {Cheng}, {de Amici}, {Fixsen}, {Hinshaw}, {Jackson}, {Janssen}, {Keegstra}, {Loewenstein}, {Lubin}, {Mather}, {Tenorio}, {Weiss}, {Wilkinson}, \& {Wright}}]{Kog1993}
{Kogut}, A., {Lineweaver}, C., {Smoot}, G.~F., {et~al.} 1993, \apj, 419, 1

\bibitem[{{Kraan-Korteweg} {et~al.}(2017){Kraan-Korteweg}, {Cluver}, {Bilicki}, {Jarrett}, {Colless}, {Elagali}, {B{\"o}hringer}, \& {Chon}}]{Kra2017}
{Kraan-Korteweg}, R.~C., {Cluver}, M.~E., {Bilicki}, M., {et~al.} 2017, \mnras, 466, L29

\bibitem[{{Libeskind} {et~al.}(2018){Libeskind}, {van de Weygaert}, {Cautun}, {Falck}, {Tempel}, {Abel}, {Alpaslan}, {Arag{\'o}n-Calvo}, {Forero-Romero}, {Gonzalez}, {Gottl{\"o}ber}, {Hahn}, {Hellwing}, {Hoffman}, {Jones}, {Kitaura}, {Knebe}, {Manti}, {Neyrinck}, {Nuza}, {Padilla}, {Platen}, {Ramachandra}, {Robotham}, {Saar}, {Shandarin}, {Steinmetz}, {Stoica}, {Sousbie}, \& {Yepes}}]{Lib2018}
{Libeskind}, N.~I., {van de Weygaert}, R., {Cautun}, M., {et~al.} 2018, \mnras, 473, 1195

\bibitem[{{Lilow} {et~al.}(2024){Lilow}, {Ganeshaiah Veena}, \& {Nusser}}]{Lil2024}
{Lilow}, R., {Ganeshaiah Veena}, P., \& {Nusser}, A. 2024, \aap, 689, A226

\bibitem[{{Macri} {et~al.}(2019){Macri}, {Kraan-Korteweg}, {Lambert}, {Alonso}, {Berlind}, {Calkins}, {Erdo{\u{g}}du}, {Falco}, {Jarrett}, \& {Mink}}]{Mac2019}
{Macri}, L.~M., {Kraan-Korteweg}, R.~C., {Lambert}, T., {et~al.} 2019, \apjs, 245, 6

\bibitem[{{Marra} {et~al.}(2013){Marra}, {Amendola}, {Sawicki}, \& {Valkenburg}}]{Mar2013}
{Marra}, V., {Amendola}, L., {Sawicki}, I., \& {Valkenburg}, W. 2013, \prl, 110, 241305

\bibitem[{{Mould} {et~al.}(2024){Mould}, {Jarrett}, {Courtois}, {Bosma}, {Deg}, {Dupuy}, {Staveley-Smith}, {Taylor}, {English}, {Rajohnson}, {Kraan-Korteweg}, {Forbes}, {D{\'e}nes}, {Lee-Waddell}, {Shen}, {Wong}, {Holwerda}, {Koribalski}, {Leahy}, {Pi{\~n}a}, \& {Yu}}]{Mou2024}
{Mould}, J., {Jarrett}, T.~H., {Courtois}, H., {et~al.} 2024, \mnras, 533, 925

\bibitem[{{Nadathur} \& {Crittenden}(2016)}]{Nad2016}
{Nadathur}, S. \& {Crittenden}, R. 2016, \apjl, 830, L19

\bibitem[{{Nusser} {et~al.}(2014){Nusser}, {Davis}, \& {Branchini}}]{Nus2014}
{Nusser}, A., {Davis}, M., \& {Branchini}, E. 2014, \apj, 788, 157

\bibitem[{{Oort}(1983)}]{Oor1983}
{Oort}, J.~H. 1983, \araa, 21, 373

\bibitem[{{Planck Collaboration} {et~al.}(2016{\natexlab{a}}){Planck Collaboration}, {Adam}, {Ade}, {Aghanim}, {Arnaud}, {Ashdown}, {Aumont}, {Baccigalupi}, {Banday}, {Barreiro}, {Bartlett}, {Bartolo}, {Basak}, {Battaner}, {Benabed}, {Beno{\^\i}t}, {Benoit-L{\'e}vy}, {Bernard}, {Bersanelli}, {Bielewicz}, {Bock}, {Bonaldi}, {Bonavera}, {Bond}, {Borrill}, {Bouchet}, {Boulanger}, {Bucher}, {Burigana}, {Butler}, {Calabrese}, {Cardoso}, {Casaponsa}, {Castex}, {Catalano}, {Challinor}, {Chamballu}, {Chary}, {Chiang}, {Christensen}, {Clements}, {Colombi}, {Colombo}, {Combet}, {Couchot}, {Coulais}, {Crill}, {Curto}, {Cuttaia}, {Danese}, {Davies}, {Davis}, {de Bernardis}, {de Rosa}, {de Zotti}, {Delabrouille}, {D{\'e}sert}, {Dickinson}, {Diego}, {Dole}, {Donzelli}, {Dor{\'e}}, {Douspis}, {Ducout}, {Dupac}, {Efstathiou}, {Elsner}, {En{\ss}lin}, {Eriksen}, {Falgarone}, {Fantaye}, {Fergusson}, {Finelli}, {Forni}, {Frailis}, {Fraisse}, {Franceschi}, {Frejsel}, {Galeotta}, {Galli}, {Ganga}, {Ghosh}, {Giard},
  {Giraud-H{\'e}raud}, {Gjerl{\o}w}, {Gonz{\'a}lez-Nuevo}, {G{\'o}rski}, {Gratton}, {Gregorio}, {Gruppuso}, {Gudmundsson}, {Hansen}, {Hanson}, {Harrison}, {Helou}, {Henrot-Versill{\'e}}, {Hern{\'a}ndez-Monteagudo}, {Herranz}, {Hildebrandt}, {Hivon}, {Hobson}, {Holmes}, {Hornstrup}, {Hovest}, {Huffenberger}, {Hurier}, {Jaffe}, {Jaffe}, {Jones}, {Juvela}, {Keih{\"a}nen}, {Keskitalo}, {Kisner}, {Kneissl}, {Knoche}, {Krachmalnicoff}, {Kunz}, {Kurki-Suonio}, {Lagache}, {Lamarre}, {Lasenby}, {Lattanzi}, {Lawrence}, {Le Jeune}, {Leonardi}, {Lesgourgues}, {Levrier}, {Liguori}, {Lilje}, {Linden-V{\o}rnle}, {L{\'o}pez-Caniego}, {Lubin}, {Mac{\'\i}as-P{\'e}rez}, {Maggio}, {Maino}, {Mandolesi}, {Mangilli}, {Maris}, {Marshall}, {Martin}, {Mart{\'\i}nez-Gonz{\'a}lez}, {Masi}, {Matarrese}, {McGehee}, {Meinhold}, {Melchiorri}, {Mendes}, {Mennella}, {Migliaccio}, {Mitra}, {Miville-Desch{\^e}nes}, {Molinari}, {Moneti}, {Montier}, {Morgante}, {Mortlock}, {Moss}, {Munshi}, {Murphy}, {Naselsky}, {Nati}, {Natoli}, {Netterfield},
  {N{\o}rgaard-Nielsen}, {Noviello}, {Novikov}, {Novikov}, {Oxborrow}, {Paci}, {Pagano}, {Pajot}, {Paladini}, {Paoletti}, {Pasian}, {Patanchon}, {Pearson}, {Perdereau}, {Perotto}, {Perrotta}, {Pettorino}, {Piacentini}, {Piat}, {Pierpaoli}, {Pietrobon}, {Plaszczynski}, {Pointecouteau}, {Polenta}, {Pratt}, {Pr{\'e}zeau}, {Prunet}, {Puget}, {Rachen}, {Racine}, {Reach}, {Rebolo}, {Reinecke}, {Remazeilles}, {Renault}, {Renzi}, {Ristorcelli}, {Rocha}, {Rosset}, {Rossetti}, {Roudier}, {Rubi{\~n}o-Mart{\'\i}n}, {Rusholme}, {Sandri}, {Santos}, {Savelainen}, {Savini}, {Scott}, {Seiffert}, {Shellard}, {Spencer}, {Stolyarov}, {Stompor}, {Sudiwala}, {Sunyaev}, {Sutton}, {Suur-Uski}, {Sygnet}, {Tauber}, {Terenzi}, {Toffolatti}, {Tomasi}, {Tristram}, {Trombetti}, {Tucci}, {Tuovinen}, {Valenziano}, {Valiviita}, {Van Tent}, {Vielva}, {Villa}, {Wade}, {Wandelt}, {Wehus}, {Yvon}, {Zacchei}, \& {Zonca}}]{Pla2016b}
{Planck Collaboration}, {Adam}, R., {Ade}, P.~A.~R., {et~al.} 2016{\natexlab{a}}, \aap, 594, A9

\bibitem[{{Planck Collaboration} {et~al.}(2016{\natexlab{b}}){Planck Collaboration}, {Ade}, {Aghanim}, {Arnaud}, {Ashdown}, {Aumont}, {Baccigalupi}, {Banday}, {Barreiro}, {Bartlett}, {Bartolo}, {Battaner}, {Battye}, {Benabed}, {Beno{\^\i}t}, {Benoit-L{\'e}vy}, {Bernard}, {Bersanelli}, {Bielewicz}, {Bock}, {Bonaldi}, {Bonavera}, {Bond}, {Borrill}, {Bouchet}, {Boulanger}, {Bucher}, {Burigana}, {Butler}, {Calabrese}, {Cardoso}, {Catalano}, {Challinor}, {Chamballu}, {Chary}, {Chiang}, {Chluba}, {Christensen}, {Church}, {Clements}, {Colombi}, {Colombo}, {Combet}, {Coulais}, {Crill}, {Curto}, {Cuttaia}, {Danese}, {Davies}, {Davis}, {de Bernardis}, {de Rosa}, {de Zotti}, {Delabrouille}, {D{\'e}sert}, {Di Valentino}, {Dickinson}, {Diego}, {Dolag}, {Dole}, {Donzelli}, {Dor{\'e}}, {Douspis}, {Ducout}, {Dunkley}, {Dupac}, {Efstathiou}, {Elsner}, {En{\ss}lin}, {Eriksen}, {Farhang}, {Fergusson}, {Finelli}, {Forni}, {Frailis}, {Fraisse}, {Franceschi}, {Frejsel}, {Galeotta}, {Galli}, {Ganga}, {Gauthier}, {Gerbino}, {Ghosh},
  {Giard}, {Giraud-H{\'e}raud}, {Giusarma}, {Gjerl{\o}w}, {Gonz{\'a}lez-Nuevo}, {G{\'o}rski}, {Gratton}, {Gregorio}, {Gruppuso}, {Gudmundsson}, {Hamann}, {Hansen}, {Hanson}, {Harrison}, {Helou}, {Henrot-Versill{\'e}}, {Hern{\'a}ndez-Monteagudo}, {Herranz}, {Hildebrandt}, {Hivon}, {Hobson}, {Holmes}, {Hornstrup}, {Hovest}, {Huang}, {Huffenberger}, {Hurier}, {Jaffe}, {Jaffe}, {Jones}, {Juvela}, {Keih{\"a}nen}, {Keskitalo}, {Kisner}, {Kneissl}, {Knoche}, {Knox}, {Kunz}, {Kurki-Suonio}, {Lagache}, {L{\"a}hteenm{\"a}ki}, {Lamarre}, {Lasenby}, {Lattanzi}, {Lawrence}, {Leahy}, {Leonardi}, {Lesgourgues}, {Levrier}, {Lewis}, {Liguori}, {Lilje}, {Linden-V{\o}rnle}, {L{\'o}pez-Caniego}, {Lubin}, {Mac{\'\i}as-P{\'e}rez}, {Maggio}, {Maino}, {Mandolesi}, {Mangilli}, {Marchini}, {Maris}, {Martin}, {Martinelli}, {Mart{\'\i}nez-Gonz{\'a}lez}, {Masi}, {Matarrese}, {McGehee}, {Meinhold}, {Melchiorri}, {Melin}, {Mendes}, {Mennella}, {Migliaccio}, {Millea}, {Mitra}, {Miville-Desch{\^e}nes}, {Moneti}, {Montier}, {Morgante},
  {Mortlock}, {Moss}, {Munshi}, {Murphy}, {Naselsky}, {Nati}, {Natoli}, {Netterfield}, {N{\o}rgaard-Nielsen}, {Noviello}, {Novikov}, {Novikov}, {Oxborrow}, {Paci}, {Pagano}, {Pajot}, {Paladini}, {Paoletti}, {Partridge}, {Pasian}, {Patanchon}, {Pearson}, {Perdereau}, {Perotto}, {Perrotta}, {Pettorino}, {Piacentini}, {Piat}, {Pierpaoli}, {Pietrobon}, {Plaszczynski}, {Pointecouteau}, {Polenta}, {Popa}, {Pratt}, {Pr{\'e}zeau}, {Prunet}, {Puget}, {Rachen}, {Reach}, {Rebolo}, {Reinecke}, {Remazeilles}, {Renault}, {Renzi}, {Ristorcelli}, {Rocha}, {Rosset}, {Rossetti}, {Roudier}, {Rouill{\'e} d'Orfeuil}, {Rowan-Robinson}, {Rubi{\~n}o-Mart{\'\i}n}, {Rusholme}, {Said}, {Salvatelli}, {Salvati}, {Sandri}, {Santos}, {Savelainen}, {Savini}, {Scott}, {Seiffert}, {Serra}, {Shellard}, {Spencer}, {Spinelli}, {Stolyarov}, {Stompor}, {Sudiwala}, {Sunyaev}, {Sutton}, {Suur-Uski}, {Sygnet}, {Tauber}, {Terenzi}, {Toffolatti}, {Tomasi}, {Tristram}, {Trombetti}, {Tucci}, {Tuovinen}, {T{\"u}rler}, {Umana}, {Valenziano}, {Valiviita},
  {Van Tent}, {Vielva}, {Villa}, {Wade}, {Wandelt}, {Wehus}, {White}, {White}, {Wilkinson}, {Yvon}, {Zacchei}, \& {Zonca}}]{Pla2016a}
{Planck Collaboration}, {Ade}, P.~A.~R., {Aghanim}, N., {et~al.} 2016{\natexlab{b}}, \aap, 594, A13

\bibitem[{{Planck Collaboration} {et~al.}(2016{\natexlab{c}}){Planck Collaboration}, {Ade}, {Aghanim}, {Arnaud}, {Ashdown}, {Aumont}, {Baccigalupi}, {Banday}, {Barreiro}, {Bartolo}, {Basak}, {Battaner}, {Benabed}, {Beno{\^\i}t}, {Benoit-L{\'e}vy}, {Bernard}, {Bersanelli}, {Bielewicz}, {Bock}, {Bonaldi}, {Bonavera}, {Bond}, {Borrill}, {Bouchet}, {Bucher}, {Burigana}, {Butler}, {Calabrese}, {Cardoso}, {Casaponsa}, {Catalano}, {Challinor}, {Chamballu}, {Chiang}, {Christensen}, {Church}, {Clements}, {Colombi}, {Colombo}, {Combet}, {Couchot}, {Coulais}, {Crill}, {Curto}, {Cuttaia}, {Danese}, {Davies}, {Davis}, {de Bernardis}, {de Rosa}, {de Zotti}, {Delabrouille}, {D{\'e}sert}, {Diego}, {Dole}, {Donzelli}, {Dor{\'e}}, {Douspis}, {Ducout}, {Dupac}, {Efstathiou}, {Elsner}, {En{\ss}lin}, {Eriksen}, {Fergusson}, {Fernandez-Cobos}, {Finelli}, {Forni}, {Frailis}, {Fraisse}, {Franceschi}, {Frejsel}, {Galeotta}, {Galli}, {Ganga}, {G{\'e}nova-Santos}, {Giard}, {Giraud-H{\'e}raud}, {Gjerl{\o}w}, {Gonz{\'a}lez-Nuevo},
  {G{\'o}rski}, {Gratton}, {Gregorio}, {Gruppuso}, {Gudmundsson}, {Hansen}, {Hanson}, {Harrison}, {Henrot-Versill{\'e}}, {Hern{\'a}ndez-Monteagudo}, {Herranz}, {Hildebrandt}, {Hivon}, {Hobson}, {Holmes}, {Hornstrup}, {Hovest}, {Huffenberger}, {Hurier}, {Ili{\'c}}, {Jaffe}, {Jaffe}, {Jones}, {Juvela}, {Keih{\"a}nen}, {Keskitalo}, {Kisner}, {Kneissl}, {Knoche}, {Kunz}, {Kurki-Suonio}, {Lagache}, {L{\"a}hteenm{\"a}ki}, {Lamarre}, {Langer}, {Lasenby}, {Lattanzi}, {Lawrence}, {Leonardi}, {Lesgourgues}, {Levrier}, {Liguori}, {Lilje}, {Linden-V{\o}rnle}, {L{\'o}pez-Caniego}, {Lubin}, {Ma}, {Mac{\'\i}as-P{\'e}rez}, {Maggio}, {Maino}, {Mandolesi}, {Mangilli}, {Marcos-Caballero}, {Maris}, {Martin}, {Mart{\'\i}nez-Gonz{\'a}lez}, {Masi}, {Matarrese}, {McGehee}, {Meinhold}, {Melchiorri}, {Mendes}, {Mennella}, {Migliaccio}, {Mitra}, {Miville-Desch{\^e}nes}, {Moneti}, {Montier}, {Morgante}, {Mortlock}, {Moss}, {Munshi}, {Murphy}, {Naselsky}, {Nati}, {Natoli}, {Netterfield}, {N{\o}rgaard-Nielsen}, {Noviello}, {Novikov},
  {Novikov}, {Oxborrow}, {Paci}, {Pagano}, {Pajot}, {Paoletti}, {Pasian}, {Patanchon}, {Perdereau}, {Perotto}, {Perrotta}, {Pettorino}, {Piacentini}, {Piat}, {Pierpaoli}, {Pietrobon}, {Plaszczynski}, {Pointecouteau}, {Polenta}, {Popa}, {Pratt}, {Pr{\'e}zeau}, {Prunet}, {Puget}, {Rachen}, {Reach}, {Rebolo}, {Reinecke}, {Remazeilles}, {Renault}, {Renzi}, {Ristorcelli}, {Rocha}, {Rosset}, {Rossetti}, {Roudier}, {Rubi{\~n}o-Mart{\'\i}n}, {Rusholme}, {Sandri}, {Santos}, {Savelainen}, {Savini}, {Schaefer}, {Scott}, {Seiffert}, {Shellard}, {Spencer}, {Stolyarov}, {Stompor}, {Sudiwala}, {Sunyaev}, {Sutton}, {Suur-Uski}, {Sygnet}, {Tauber}, {Terenzi}, {Toffolatti}, {Tomasi}, {Tristram}, {Tucci}, {Tuovinen}, {Valenziano}, {Valiviita}, {Van Tent}, {Vielva}, {Villa}, {Wade}, {Wandelt}, {Wehus}, {Yvon}, {Zacchei}, \& {Zonca}}]{Pla2016c}
{Planck Collaboration}, {Ade}, P.~A.~R., {Aghanim}, N., {et~al.} 2016{\natexlab{c}}, \aap, 594, A21

\bibitem[{{Plionis} \& {Kolokotronis}(1998)}]{Pli1998}
{Plionis}, M. \& {Kolokotronis}, V. 1998, \apj, 500, 1

\bibitem[{{Plionis} \& {Valdarnini}(1991)}]{Pli1991}
{Plionis}, M. \& {Valdarnini}, R. 1991, \mnras, 249, 46

\bibitem[{{Pratt} {et~al.}(2009){Pratt}, {Croston}, {Arnaud}, \& {B{\"o}hringer}}]{Pra2009}
{Pratt}, G.~W., {Croston}, J.~H., {Arnaud}, M., \& {B{\"o}hringer}, H. 2009, \aap, 498, 361

\bibitem[{{Rajohnson} {et~al.}(2024{\natexlab{a}}){Rajohnson}, {Kraan-Korteweg}, {Chen}, {Frank}, {Steyn}, {Kurapati}, {Pisano}, {Staveley-Smith}, {Serra}, {Goedhart}, \& {Camilo}}]{Raj2024a}
{Rajohnson}, S. H.~A., {Kraan-Korteweg}, R.~C., {Chen}, H., {et~al.} 2024{\natexlab{a}}, \mnras, 531, 3486

\bibitem[{{Rajohnson} {et~al.}(2024{\natexlab{b}}){Rajohnson}, {Kraan-Korteweg}, {Frank}, {Chen}, {Staveley-Smith}, {Serra}, {Steyn}, {Kurapati}, {Pisano}, \& {Goedhart}}]{Raj2024b}
{Rajohnson}, S. H.~A., {Kraan-Korteweg}, R.~C., {Frank}, B.~S., {et~al.} 2024{\natexlab{b}}, \mnras, 535, 3429

\bibitem[{{Reisenegger} {et~al.}(2000){Reisenegger}, {Quintana}, {Carrasco}, \& {Maze}}]{Rei2000}
{Reisenegger}, A., {Quintana}, H., {Carrasco}, E.~R., \& {Maze}, J. 2000, \aj, 120, 523

\bibitem[{{Rowan-Robinson} {et~al.}(2000){Rowan-Robinson}, {Sharpe}, {Oliver}, {Keeble}, {Canavezes}, {Saunders}, {Taylor}, {Valentine}, {Frenk}, {Efstathiou}, {McMahon}, {White}, {Sutherland}, {Tadros}, \& {Maddox}}]{Row2000}
{Rowan-Robinson}, M., {Sharpe}, J., {Oliver}, S.~J., {et~al.} 2000, \mnras, 314, 375

\bibitem[{{Sachs} \& {Wolfe}(1967)}]{Sac1967}
{Sachs}, R.~K. \& {Wolfe}, A.~M. 1967, in Liege International Astrophysical Colloquia, Vol.~15, Liege International Astrophysical Colloquia, 59

\bibitem[{{Scaramella} {et~al.}(1991){Scaramella}, {Vettolani}, \& {Zamorani}}]{Sca1991}
{Scaramella}, R., {Vettolani}, G., \& {Zamorani}, G. 1991, \apjl, 376, L1

\bibitem[{{Springel} {et~al.}(2005){Springel}, {White}, {Jenkins}, {Frenk}, {Yoshida}, {Gao}, {Navarro}, {Thacker}, {Croton}, {Helly}, {Peacock}, {Cole}, {Thomas}, {Couchman}, {Evrard}, {Colberg}, \& {Pearce}}]{Spr2005}
{Springel}, V., {White}, S. D.~M., {Jenkins}, A., {et~al.} 2005, \nat, 435, 629

\bibitem[{{Strauss} {et~al.}(1992){Strauss}, {Yahil}, {Davis}, {Huchra}, \& {Fisher}}]{Str1992}
{Strauss}, M.~A., {Yahil}, A., {Davis}, M., {Huchra}, J.~P., \& {Fisher}, K. 1992, \apj, 397, 395

\bibitem[{{Tinker} {et~al.}(2010){Tinker}, {Robertson}, {Kravtsov}, {Klypin}, {Warren}, {Yepes}, \& {Gottl{\"o}ber}}]{Tin2010}
{Tinker}, J.~L., {Robertson}, B.~E., {Kravtsov}, A.~V., {et~al.} 2010, \apj, 724, 878

\bibitem[{{Truemper}(1993)}]{Tru1993}
{Truemper}, J. 1993, Science, 260, 1769

\bibitem[{{Tully} {et~al.}(2016){Tully}, {Courtois}, \& {Sorce}}]{Tul2016}
{Tully}, R.~B., {Courtois}, H.~M., \& {Sorce}, J.~G. 2016, \aj, 152, 50

\bibitem[{{Tully} {et~al.}(2019){Tully}, {Pomar{\`e}de}, {Graziani}, {Courtois}, {Hoffman}, \& {Shaya}}]{Tul2019}
{Tully}, R.~B., {Pomar{\`e}de}, D., {Graziani}, R., {et~al.} 2019, \apj, 880, 24

\bibitem[{{Whitbourn} \& {Shanks}(2014)}]{Whi2014}
{Whitbourn}, J.~R. \& {Shanks}, T. 2014, \mnras, 437, 2146

\bibitem[{{Yahil} {et~al.}(1980){Yahil}, {Sandage}, \& {Tammann}}]{Yah1980}
{Yahil}, A., {Sandage}, A., \& {Tammann}, G.~A. 1980, \apj, 242, 448

\end{thebibliography}
% ==================================

\begin{appendix}

\section{Cluster catalogue}

\begin{table}[ht]% Do NOT use \begin{table*}
	\caption{Catalogue of the 185 clusters in the five superstructures.
}
\medskip
\label{table:cluscat} % give each table a logical label name
\begin{tabular}{lrrrrr}\hline
\small  name & {\rm RA} & {\rm DEC} & {\rm redshift} & $L_{X,500}$ & $M_{200}$\\ 
\hline
{\rm Quipu}  &&&&& \\
\hline
RXCJ0150.7+3305 &  27.6789 &  33.0851 & 0.0347 &   0.133 &   1.056 \\
RXCJ0214.2+5144 &  33.5702 &  51.7473 & 0.0489 &   0.632 &   2.754 \\
RXCJ0228.1+2811 &  37.0413 &  28.1940 & 0.0359 &   0.183 &   1.287 \\
RXCJ0229.0+3805 &  37.2543 &  38.0964 & 0.0382 &   0.143 &   1.001 \\
RXCJ0229.9+2307 &  37.4793 &  23.1172 & 0.0307 &   0.107 &   0.922 \\
RXCJ0246.0+3653 &  41.5149 &  36.8865 & 0.0473 &   0.686 &   2.902 \\
RXCJ0251.1+4513 &  42.7979 &  45.2237 & 0.0440 &   0.095 &   0.854 \\
RXCJ0254.0+3625 &  43.5042 &  36.4294 & 0.0474 &   0.120 &   0.983 \\
RXCJ0257.6+1605 &  44.4088 &  16.0932 & 0.0316 &   0.079 &   0.765 \\
RXCJ0301.8+3549 &  45.4632 &  35.8268 & 0.0462 &   0.298 &   1.729 \\
\hline
\end{tabular}	
\smallskip

{\bf Notes:} Only the first ten lines of the catalogue are shown, the full table
is only available in electronic form at CDS via anonymous ftp to cdsarc.u-strasbg.fr (130.79.128.5) or via http://cdsweb.u-strasbg.fr/cgi-bin/qcat?J/A+A/..
The X-ray luminosity, $L_{X,500}$, is in units of $10^{44}$ erg s$^{-1}$ in the 0.1 to 2.4 keV energy band and the estimated mass, $M_{200}$, is in units of $10^{14}$M$_{\odot}$.
\end{table}
    
\end{appendix}
\end{document}